\documentstyle[epsfig,longtable,color]{mn}

\newif\ifAMStwofonts

\ifoldfss
  \newcommand{\rmn}[1] {{\rm #1}}

  \renewcommand{\chem}[2] {$\rm{}^{#2}\kern-0.8pt#1$}
  \ifCUPmtlplainloaded \else
    \NewTextAlphabet{textbfit} {cmbxti10} {}
    \NewTextAlphabet{textbfss} {cmssbx10} {}
    \NewMathAlphabet{mathbfit} {cmbxti10} {} 
    \NewMathAlphabet{mathbfss} {cmssbx10} {} 
  \fi
  \ifAMStwofonts
    \ifCUPmtlplainloaded \else
      \NewSymbolFont{upmath} {eurm10}
      \NewSymbolFont{AMSa} {msam10}
      \NewMathSymbol{\upi}     {0}{upmath}{19}
      \NewMathSymbol{\umu}     {0}{upmath}{16}
      \NewMathSymbol{\upartial}{0}{upmath}{40}
      \NewMathSymbol{\leqslant}{3}{AMSa}{36}
      \NewMathSymbol{\geqslant}{3}{AMSa}{3E}

       \let\le=\leqslant
       \let\ge=\geqslant
    \fi
  \fi
\fi 

\ifnfssone
  \newmathalphabet{\mathit}
  \addtoversion{normal}{\mathit}{cmr}{m}{it}
  \addtoversion{bold}{\mathit}{cmr}{bx}{it}
  \newcommand{\rmn}[1] {\mathrm{#1}}

 \newmathalphabet{\mathbfit} 
  \addtoversion{normal}{\mathbfit}{cmr}{bx}{it}
  \addtoversion{bold}{\mathbfit}{cmr}{bx}{it}
  \newmathalphabet{\mathbfss} 
  \addtoversion{normal}{\mathbfss}{cmss}{bx}{n}
  \addtoversion{bold}{\mathbfss}{cmss}{bx}{n}
  \ifAMStwofonts
    \ifCUPmtlplainloaded \else
      \UseAMStwoboldmath
      \makeatletter
      \new@mathgroup\upmath@group
      \define@mathgroup\mv@normal\upmath@group{eur}{m}{n}
      \define@mathgroup\mv@bold\upmath@group{eur}{b}{n}
      \edef\UPM{\hexnumber\upmath@group}
      \new@mathgroup\amsa@group
      \define@mathgroup\mv@normal\amsa@group{msa}{m}{n}
      \define@mathgroup\mv@bold\amsa@group{msa}{m}{n}
      \edef\AMSa{\hexnumber\amsa@group}
      \makeatother
      \mathchardef\upi="0\UPM19
      \mathchardef\umu="0\UPM16
      \mathchardef\upartial="0\UPM40
      \mathchardef\leqslant="3\AMSa36
      \mathchardef\geqslant="3\AMSa3E

       \let\le=\leqslant
       \let\ge=\geqslant
    \fi
  \fi
\fi 

\ifnfsstwo
  \newcommand{\rmn}[1] {\mathrm{#1}}

  \DeclareMathAlphabet{\mathbfit}{OT1}{cmr}{bx}{it}
  \SetMathAlphabet\mathbfit{bold}{OT1}{cmr}{bx}{it}
  \DeclareMathAlphabet{\mathbfss}{OT1}{cmss}{bx}{n}
  \SetMathAlphabet\mathbfss{bold}{OT1}{cmss}{bx}{n}
  \ifAMStwofonts
    \ifCUPmtlplainloaded \else
      \DeclareSymbolFont{UPM}{U}{eur}{m}{n}
      \SetSymbolFont{UPM}{bold}{U}{eur}{b}{n}
      \DeclareSymbolFont{AMSa}{U}{msa}{m}{n}
      \DeclareMathSymbol{\upi}{0}{UPM}{"19}
      \DeclareMathSymbol{\umu}{0}{UPM}{"16}
      \DeclareMathSymbol{\upartial}{0}{UPM}{"40}
      \DeclareMathSymbol{\leqslant}{3}{AMSa}{"36}
      \DeclareMathSymbol{\geqslant}{3}{AMSa}{"3E}

       \let\le=\leqslant
       \let\ge=\geqslant
    \fi
  \fi
\fi 

\ifCUPmtlplainloaded \else
  \ifAMStwofonts \else 
    \def\upi{\pi}
    \def\umu{\mu}
    \def\upartial{\partial}
  \fi
\fi

\title{The \mbox{\boldmath{${\rm [OII]\lambda3727/H\alpha}$}} Ratio 
       of Emission Line Galaxies in the 2dF Galaxy Redshift Survey}
\author[Mouhcine et al.]
       {M. Mouhcine$^{1,2}$, 
        I. Lewis$^3$,  
        B. Jones$^{1,4}$,
        F. Lamareille$^5$,
        S.J. Maddox$^1$, 
        T. Contini$^5$\\
       $^1$ School of Physics and Astronomy, University of Nottingham, 
       Nottingham NG7 2RD\\
       $^2$ Observatoire Astronomique de Strasbourg (UMR 7550),
       11, rue de l'Universit\'e, 67000 Strasbourg, France \\
       $^3$ Astrophysics, Nuclear and Astrophysics Laboratory, Keble Road, 
       Oxford OX1 3RH \\
       $^4$ Astronomy Unit, School of Mathematical Sciences, Queen Mary,
       University of London, Mile End Road, London, E1 4NS \\ 
       $^5$ Laboratorie d'Astrophysique de Toulouse et Tarbes 
       (LA2T - UMR 5572), Observatoire Midi-Pyr\'en\'ees, \\
       14 Avenue E. Belin, F-31400 Toulouse, France 
        }    
\date{Accepted ?.
      Received ?;
      in original form ?}

\pubyear{2001}

\begin{document}

\maketitle

\label{firstpage}

\begin{abstract}

We investigate the systematic variation of the ${\rm
[OII]\lambda3727/H\alpha}$ flux line ratio as a function of various 
galaxy properties, i.e., luminosity, metallicity, reddening, and 
excitation state, for a sample of $1\,124$ emission-line galaxies, 
with a mean redshift $z\sim 0.06$, drawn from the Two Degree Field 
Galaxy Redshift Survey.
The mean observed and extinction-corrected emission-line flux ratios 
agree well with the values derived from the $B$-band selected Nearby 
Field Galaxy Survey galaxy sample, but are significantly different 
from the values obtained from the ${\rm H\alpha}$-selected Universidad 
Complutense de Madrid Survey galaxy sample. This is because the different 
selection criteria applied in these surveys lead to a significant 
difference in the mean extinction and metallicity of different samples.

We use the ${\rm R_{23}}$ parameter to estimate the gas-phase oxygen
abundance and find that the extinction-corrected 
${\rm [OII]\lambda3727/H\alpha}$ ratio depends on the oxygen abundance. 
For ${\rm 12+\log(O/H)\ga\,8.4}$, we confirm that the emission-line ratio 
decreases with increasing metallicity. We have extended the relationship 
further to the metal-poor regime, ${\rm 12+\log(O/H)\la 8.4}$, and find 
that the correlation between the extinction-corrected ${\rm 
[OII]\lambda3727/H\alpha}$ ratio and the metallicity reverses in comparison 
to the relationship for metal-rich galaxies. For metal-poor galaxies, 
in contrast with metal-rich ones, the variation of extinction-corrected 
${\rm [OII]\lambda3727/H\alpha}$ ratio is correlated with the ionization 
states of the interstellar gas.

The relative importance of the metallicity or the excitation state in
determining the extinction-corrected ${\rm [OII]\lambda3727/H\alpha}$
ratio depends on galaxy luminosity.

\end{abstract}

\begin{keywords}
surveys, galaxies: fundamental parameters, galaxies: general,
galaxies: statistics
\end{keywords}

\section{Introduction}
\label{intro.sec}

Spectral features in integrated spectra of galaxies allow us to
determine important aspects of their evolutionary state. With the
advent of the current generation of large telescopes,
spectrophotometric studies are possible for fainter and more distant
galaxies.  Measuring the evolution of the star formation rate since
the earliest cosmic epochs in the Universe is crucial for an accurate
understanding of the formation and evolution of galaxies. Since the
pioneering papers by Lilly et al. (1996) and Madau et al. (1996),
several deep spectroscopic surveys have enabled detailed investigations 
of the star formation history of the universe (e.g., Hammer et al. 1997, 
Tresse et al. 2002, Hippelein et al. 2003).

The interpretation of integrated spectral properties, and estimates 
of the cosmic star formation rate over an extended redshift range
requires the use of a range of star formation indicators. 
Unfortunately, there are significant discrepancies between different 
star formation rate indicators (Hopkins et al. 2003, and reference 
therein). The flux of the ${\rm H\alpha}$ Balmer line is directly 
linked to the total ionizing flux, making this line the most robust 
and reliable tracer of star formation. ${\rm H\alpha}$ emission 
line still suffers from attenuation by dust however.
The ${\rm [OII]\lambda3727}$ emission line has been used widely in 
a number of studies of the star formation rate in redshift ranges 
where the ${\rm H\alpha}$ emission line moves into the near-infrared 
(e.g., Thompson \& Djorgovski 1991; Cowie et al. 1997; Hogg et al. 
1998; Hippelein et al. 2003). However, published calibrations of 
the star formation rate in terms of the ${\rm [OII]\lambda3727}$ 
emission vary by factors of a few (Gallagher et al. 1989; Kennicutt 
1992; Guzm\'an et al. 1997; Rosa-Gonz\`alez et al. 2002). 

Jansen et al. (2001) and Kewley et al. (2004) used the Nearby Field
Galaxy Survey (Jansen et al. 2000, NFGS hereafter), and
Arag\'on-Salamanca et al.  (2005, submitted) the ${\rm H\alpha}$
selected Universidad Complutense de Madrid Survey (UCM) to investigate the 
variation of ${\rm [OII]\lambda3727}$ as a function of galaxy properties. 
Kewley et al. (2004) have found no systematic difference between star 
formation rates using ${\rm H\alpha}$ and ${\rm [OII]\lambda3727}$ 
luminosities after correcting for the effects of internal extinction 
and metallicity on ${\rm [OII]\lambda3727}$ luminosity. Unfortunately, 
these are relatively small surveys and they sample rather limited 
ranges of interstellar gas parameters.

Since the pioneering works by Gallagher et al. (1989) and Kennicutt 
(1992), several spectroscopic studies of galaxies in the Local Universe 
have been undertaken (e.g., Tresse et al. 1999, Salzer et al. 2000, 
Jansen et al 2000, Carter et al. 2001, Gavazzi et al. 2004). However, 
the number of emission line galaxies observed in these surveys for 
which {\it all} the emission lines needed to identify the nature of 
the ionizing source {\it and} to estimate gas phase metallicity 
(i.e., ${\rm [OII]\lambda3727}$, ${\rm H\beta}$, ${\rm 
[OIII]\lambda4959,\lambda5007}$, ${\rm [NII]\lambda6548}$, ${\rm 
H\alpha}$, ${\rm [NII]\lambda6584}$, ${\rm [SII]\lambda6717}$, and 
${\rm [SII]\lambda6731}$) are observed with high confidence do not 
exceed a few hundred galaxies at best. 
The advent of large spectroscopic surveys, such as the Sloan Digital 
Sky Survey (Stoughton et al. 2002, Abazaijan et al 2003, SDSS hereafter) 
and Two Degree Field Galaxy Redshift Survey (Colless et al. 2001, 
2dFGRS hereafter) provides larger emission line galaxy samples with 
all the needed emission lines.

The 2dFGRS was carried out with the primary aim of studying the
three-dimensional clustering properties of galaxies and determining
the luminosity function. However for a subsample of the 2dFGRS 
galaxies, the quality of the spectra is good enough to determine 
accurate emission line properties. We select normal star forming 
emission line galaxies with strong emission lines, high-quality 
spectra, and high signal-to-noise for a detailed study of the 
sensitivity of ${\rm [OII]\lambda3727/H\alpha}$ flux ratio to galaxy
and interstellar medium properties. We aim to establish the properties 
of a local sample that can be used as a comparison for more distant 
galaxy samples. The 2dFGRS spectra are suitable for carrying out such 
an investigation, and have the following properties: (i) the large 
spectral coverage means that the galaxy spectra contain most of the 
prominent optical emission lines, including ${\rm [OII]\lambda3727}$ 
and ${\rm H\alpha}$, and emission lines needed to identify the ionizing 
source, (ii) we can correct ${\rm H\alpha}$ and ${\rm H\beta}$ for the 
absorption features in the spectrum of the underlying stellar population.

The paper is organized as follows. In Sect.\ref{data.sec}, we describe
how we obtain the emission line galaxy sample used in this paper from
the original 2dFGRS data set. In Section \ref{aper}, we describe
surface photometry measurements of the 2dFGRS galaxies, and assess
the effects of aperture on the emission line properties. 
Section \ref{oii_ha} discusses the dependence of emission line 
$\rm [OII]\lambda3727/{\rm H\alpha}$ flux ratio on interstellar gas 
properties.  In Sect. \ref{concl.sec}, we present the results of this 
analysis and summarize our conclusions. 

Throughout this paper, all calculations assume the cosmology given
by \textit{WMAP}, with $\Omega_{\Lambda}=0.73$, $\Omega_{m}=0.27$
and $\rmn{H}_{0}=71\rmn{km}\ \rmn{s}^{-1}\ \rmn{Mpc}^{-1}$ 
(Spergel et al. 2003).

\section{Galaxy sample}
\label{data.sec}

Our sample is drawn from the 2dFGRS data set, which consists of 
optical (3600-8000 {\AA}) spectroscopy of more than 250\,000 galaxies 
brighter than $b_j=19.7$, with a full width at half-maximum (FWHM 
hereafter) spectral resolution of 9 {\AA}. The survey covers two 
contiguous declination strips, plus 99 randomly located fields. 
One of the strips is located close to the south Galactic pole, 
while the other strip is located on the celestial equator in the 
northern Galactic hemisphere. Full details of the survey strategy 
are given in Colless et al. (2001).

We first exclude all galaxies observed before 31 August 1999, since
these were observed whilst there was a fault with the atmospheric
dispersion compensator within the 2dF instrument (Lewis et al. 2002a). 
For these galaxies, the fitting procedure to determine the line 
properties gives results of poor quality. This cut leaves us with 
200\,160 galaxies. We selected galaxies with high quality redshift 
determinations, reducing the sample to 185\,731 galaxies. 
We used a fully automatic procedure to measure the emission lines
properties. A detailed discussion of the procedure and the 
determination of the fitting quality is presented by Lewis et al. 
(2002b). Here we summarize the basic points of this procedure.
The line fitting consists of a simultaneous fitting of a series of 
absorption lines and a series of emission lines. Some of these are 
very close in wavelength (e.g., ${\rm H\beta}$ in absorption and 
emission). This technique works very well in fitting broad absorption 
lines and narrow emission lines (see figure $2$ in Lewis et al, 2002b).
The fitting allows a common wavelength shift for all the lines so 
relative shifting between the lines is not allowed. It does not work 
as well for broad emission lines where the absorption component is 
not well constrained or where several emission lines can combine 
to give a non-unique solution (e.g., ${\rm [NII]\lambda6548}$ and 
${\rm H\alpha}$). These results are however, good enough to identify 
broad emission line cases. Note that only high signal-to-noise (S/N)
spectra have been fitted. 
To get accurate estimates of the gas phase properties, and to avoid 
any possible bias, we select galaxies having a good quality fit for 
{\it all} of the emission lines needed to classify the galaxies, i.e., 
the nature of the ionizing sources, and to measure the gas-phase oxygen 
abundance. This requirement leaves us with a sample of 10\,284 galaxies. 
Equivalent widths were corrected from the observed to the rest frame. 
We select only galaxies whose spectra have a relatively high S/N ratio, 
i.e. ${\rm S/N\ge 10}$ measured on the continuum between 4000 {\AA}
and 7500 {\AA}. 
This leaves us with a sample of 7\,353 galaxies. In our subsequent 
analyses, we use only galaxies which show Balmer lines in emission 
with equivalent widths larger than 10\,{\AA} after correcting 
for the underlying stellar absorption. This corresponds to the spectral 
resolution of the 2dF instrument, and galaxies with weaker emission 
are subject to large systematic uncertainties from instrumental effects, 
particularly affecting estimates of the internal dust extinction from 
the Balmer decrement. This equivalent width threshold also minimizes 
the effect of the underlying stellar absorption. The requirement of 
having the ${\rm H\beta}$ equivalent width larger than the spectral 
resolution of the 2dF instrument drastically reduces the number of 
emission line galaxies in the final sample, and leaves us with 
$1\,327$ galaxies. 

\begin{figure*}
\includegraphics[clip=,width=0.45\textwidth]{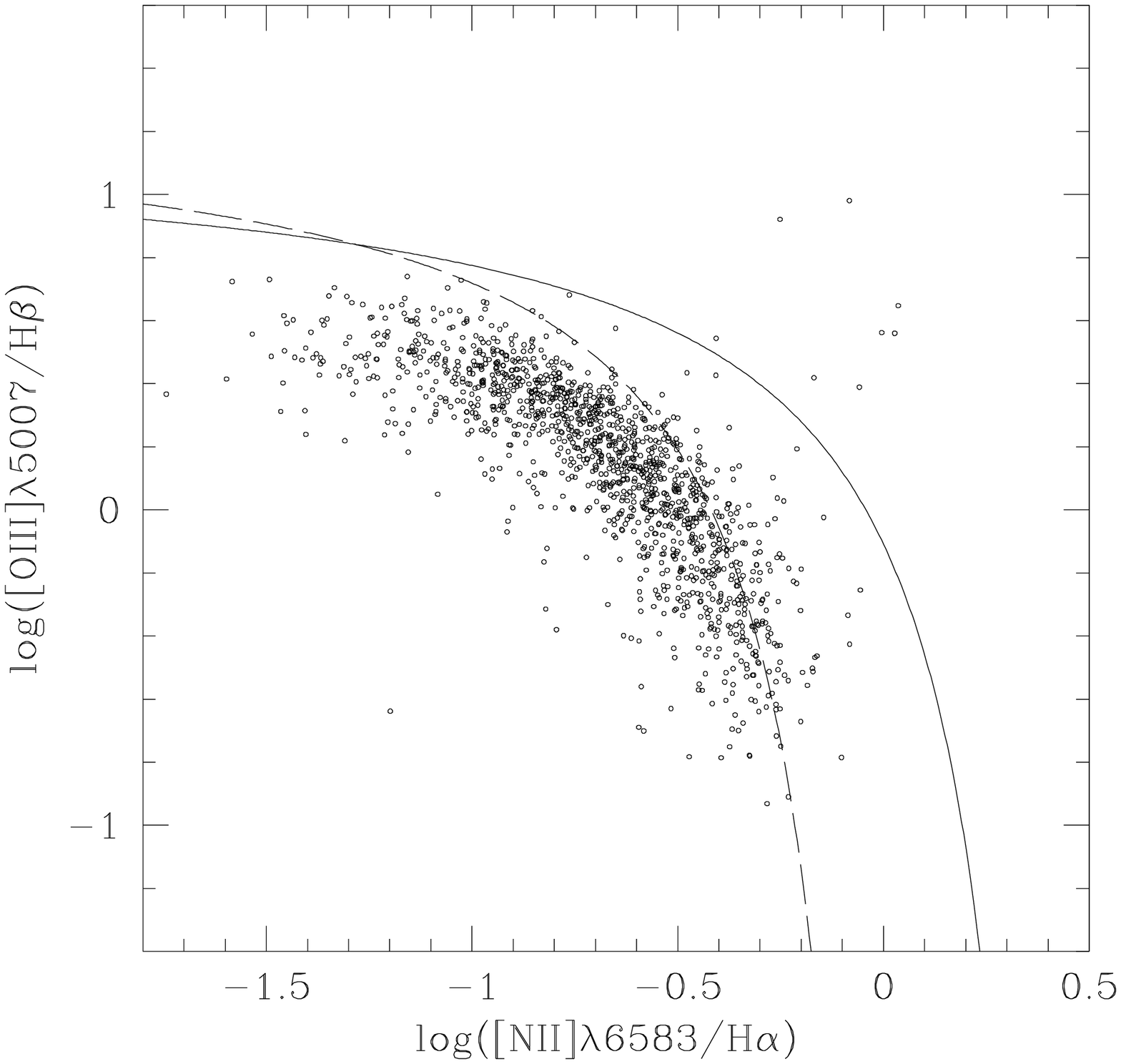}
\includegraphics[clip=,width=0.45\textwidth]{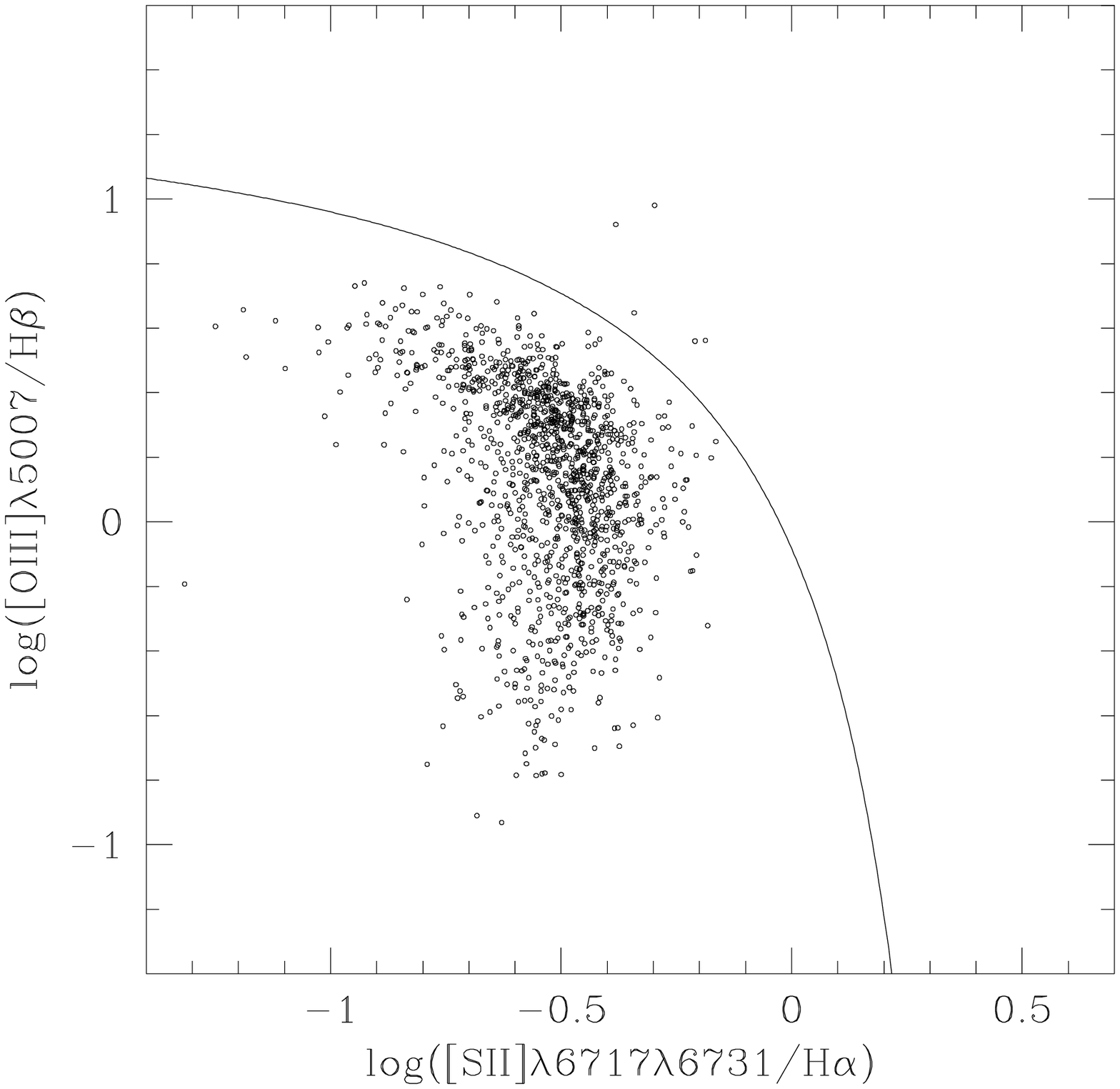}
\caption{Diagnostic diagrams for our sample of 1327 narrow emission line 
galaxies. The continuous lines show the theoretical separation between 
starburst galaxies and AGNs from Kewley et al. (2001). The dashed line in 
the ${\rm [NII]\lambda6583/H\alpha}$ vs. ${\rm [OIII]\lambda5007/H\beta}$
diagram shows the separation between starburst galaxies and AGNs as 
defined empirically by Kauffmann et al. (2003) using SDSS data.}
\label{diagn_diagr}
\end{figure*}

As we are interested in normal emission line galaxies, we have excluded 
galaxies which are dominated by Active Galactic Nuclei (AGN hereafter).
We first exclude galaxies which have a ${\rm H\beta}$ emission line 
FWHM larger than 10\,{\AA} (corresponding to a velocity width of 
$\ga 670$ kms$^{-1}$), 
since these are likely to be Seyfert I galaxies. 
We then use the classical diagnostic ratios of two pairs 
of relatively strong emission lines (Baldwin et al. 1981, Veilleux 
\& Osterbrock 1987) to distinguish between galaxies dominated by 
emission from star-forming regions and galaxies dominated by emission 
from non-thermal ionizing sources. We classify galaxies according to
their position in ${\rm [OIII]\lambda5007/H\beta}$ vs. ${\rm 
[NII]\lambda6583/H\alpha}$ and ${\rm [OIII]\lambda5007/H\beta}$ vs. 
${\rm [SII]\lambda6717,\lambda6731/H\alpha}$ diagrams. The demarcation
between star-forming galaxies and AGN in both diagrams was taken from
Kewley et al. (2001). Fig.\,\ref{diagn_diagr} shows the distribution of 
the sample galaxies in the diagnostic diagrams. We used the conservative 
requirement that a galaxy must be classified as a star-forming galaxy 
in both diagnostic diagrams in order to be retained in our sample (see 
Lamareille et al. 2004 for more detail on the classification of emission 
line objects). The diagrams show that our sample contains galaxies with 
a large range of excitation levels, suggesting that our sample contains 
both metal-rich and metal-poor galaxies. This sample is thus suitable 
for studying the properties of dust obscuration and emission lines 
over a large range of metallicities and excitation parameters.

$196$ galaxies show Balmer decrements smaller than the intrinsic 
${\rm H\alpha/H\beta}$ ratio of 2.85 which corresponds to case B 
recombination with a temperature of ${\rm T=10^4 K}$, and a density 
of ${\rm n_{e}\sim\,10^2-10^4\,cm^{-2}}$ (Osterbrock 1989). This is 
probably due to an intrinsically low extinction, coupled with 
uncertainties in the correction of the underlying stellar absorption, 
and/or errors in the data reduction. As this implies a physically 
impossible negative extinction, those galaxies were removed from the 
sample. Thus, we end up with a final sample of 1\,124 normal emission 
line galaxies. 

\begin{figure}
\includegraphics[clip=,width=0.2\textwidth]{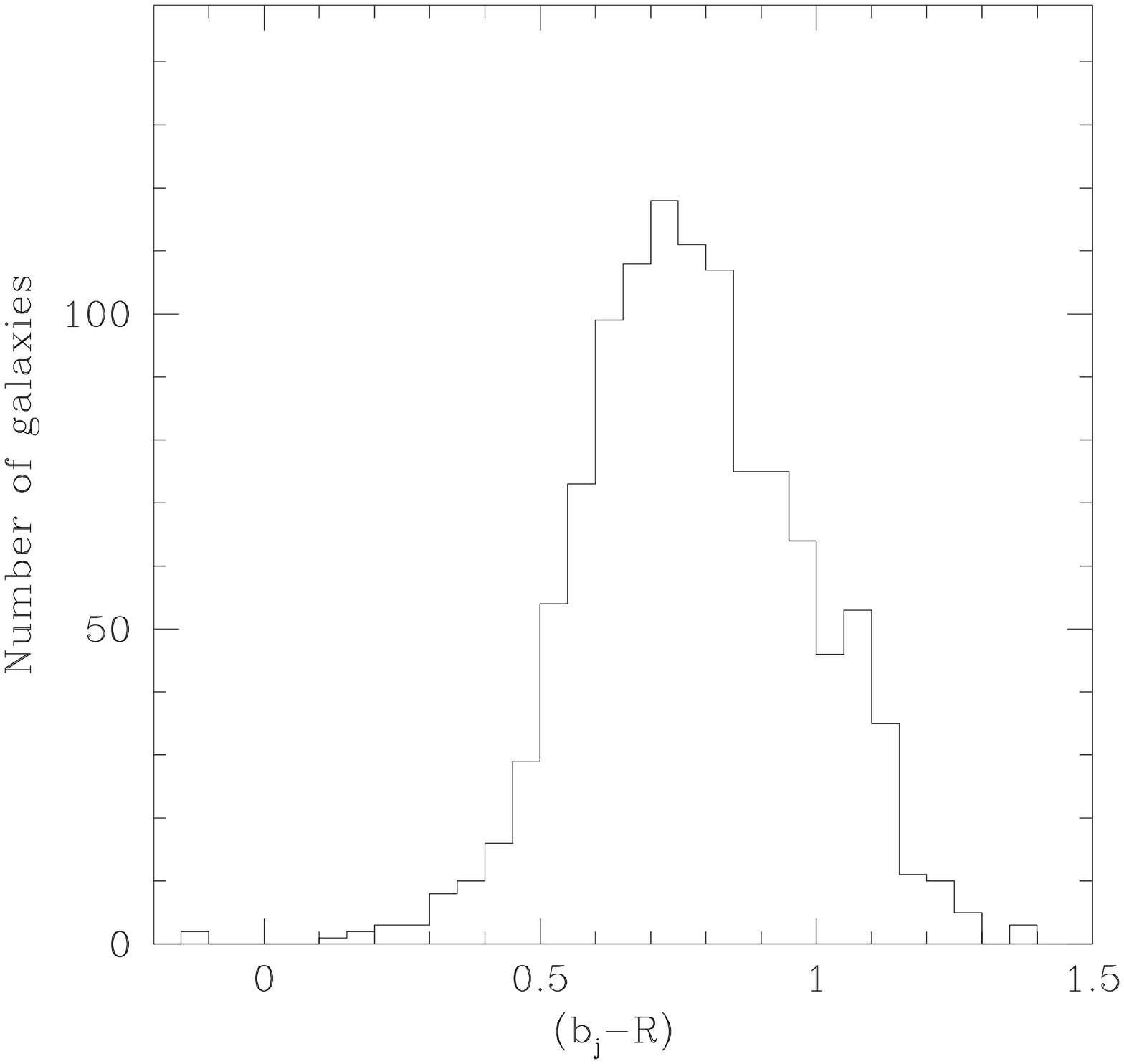}
\includegraphics[clip=,width=0.2\textwidth]{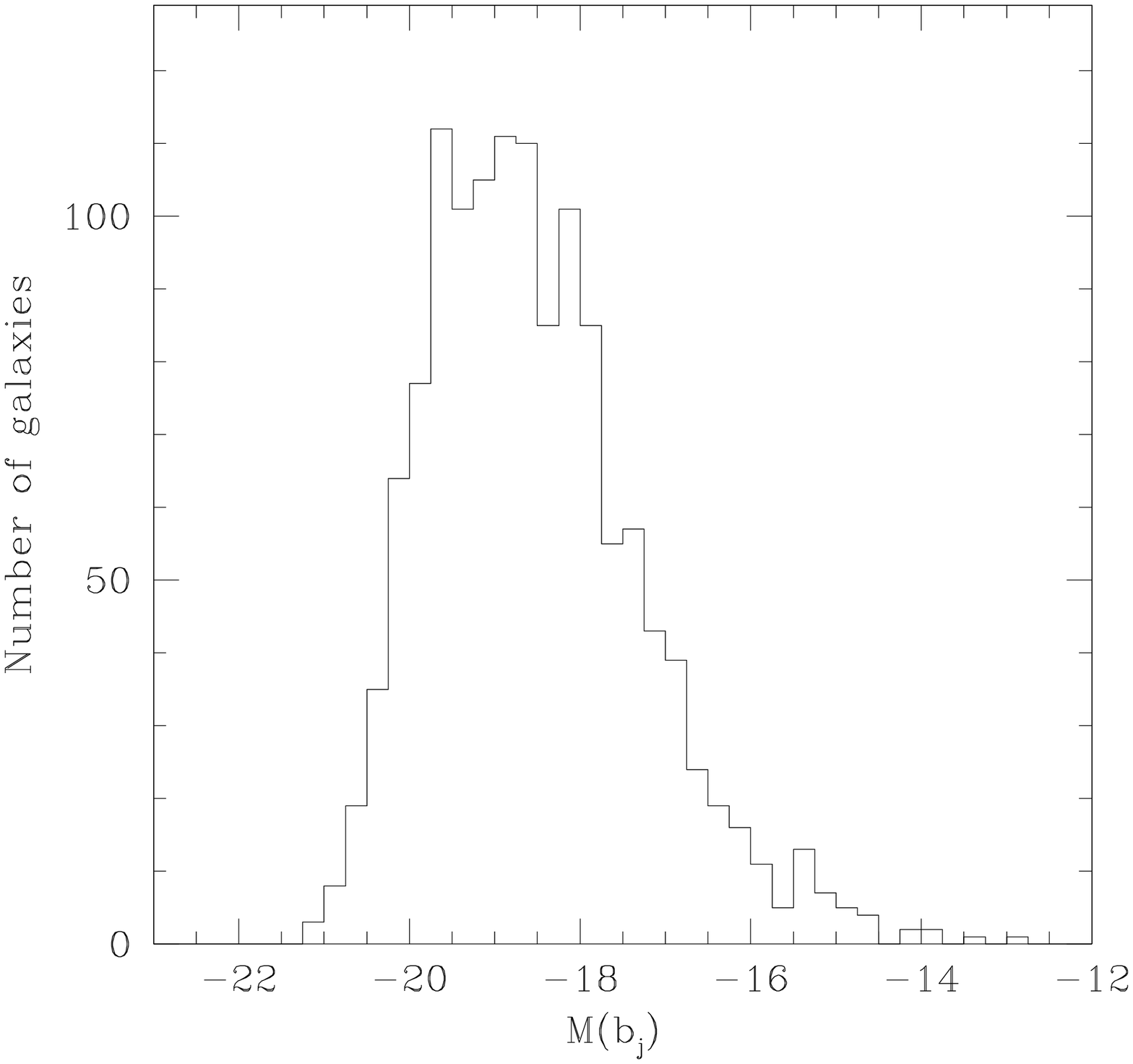}
\includegraphics[clip=,width=0.2\textwidth]{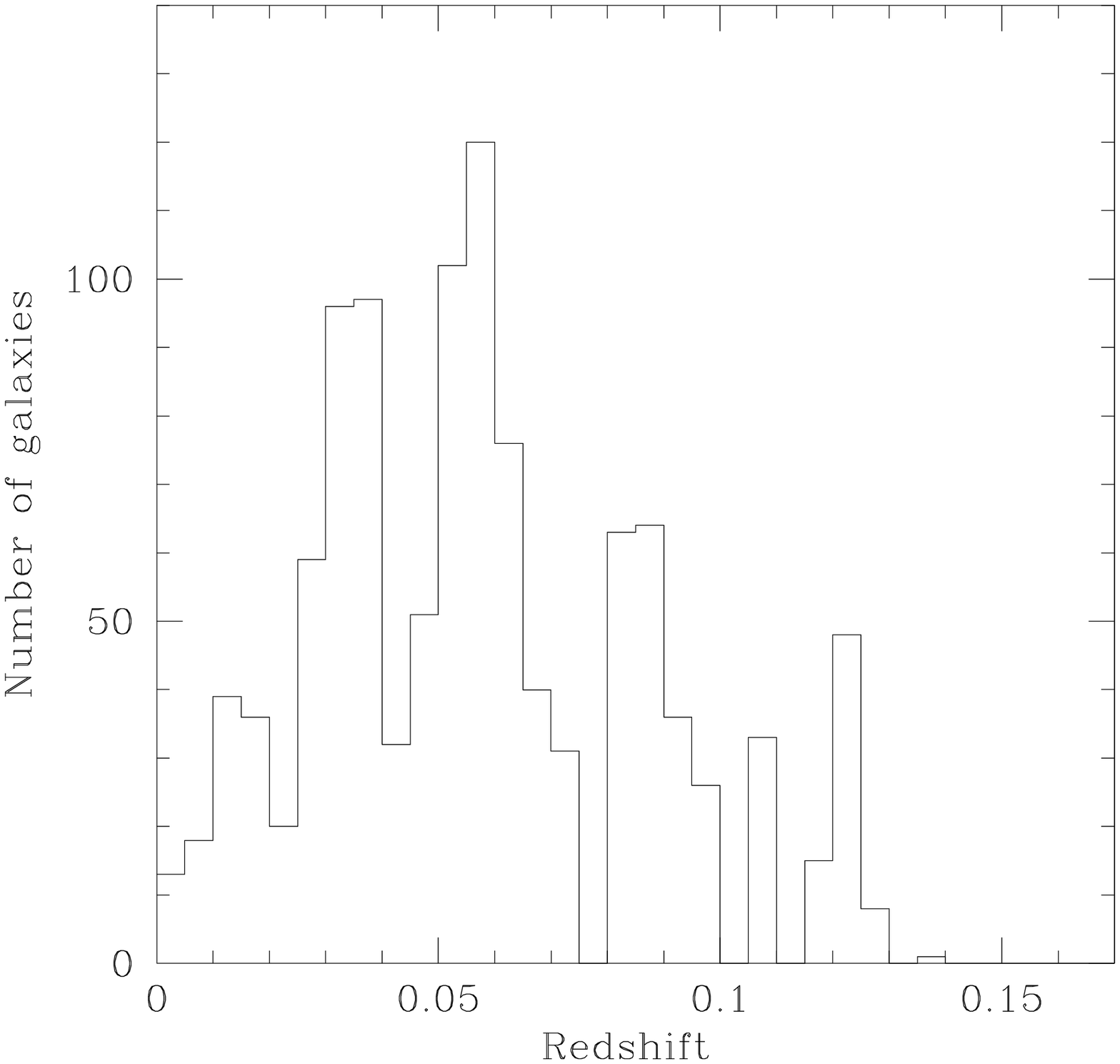}
\includegraphics[clip=,width=0.2\textwidth]{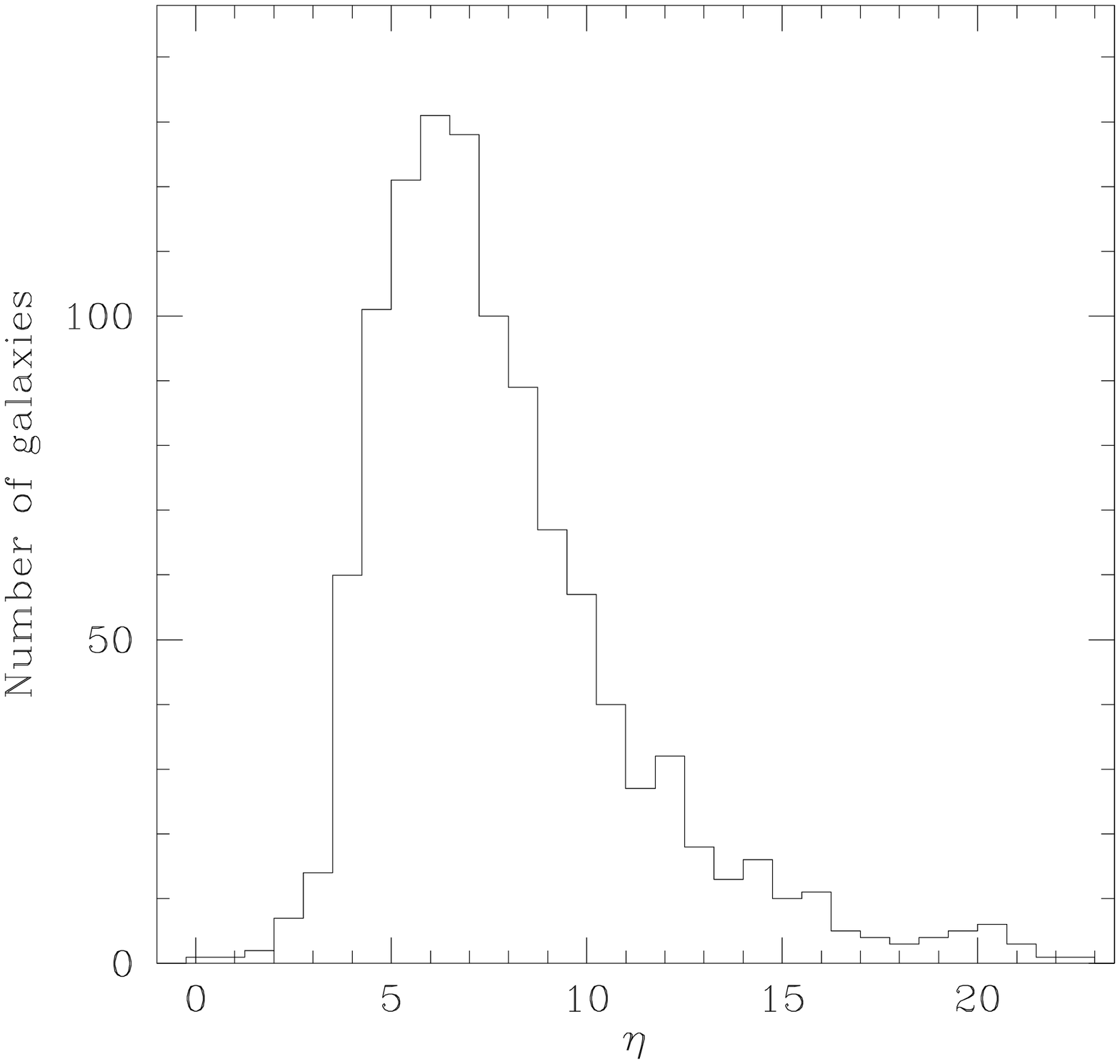}
\caption{Distribution of general properties of our sample of 1\,124 
star forming galaxies. The upper panels show the absolute $b_j$-band 
magnitude, and ($b_j$-R) colour distributions. The lower panels show 
the distributions of redshift and the spectral type-sensitive parameter 
$\eta$.}
\label{dist_sample}
\end{figure}

The distributions of global properties, i.e., galaxy colours, $b_j$-band
absolute magnitudes, redshift, and the $\eta$ parameter respectively, 
of the selected sample are shown in Fig.~\ref{dist_sample}. 
The corresponding numerical data for emission line galaxy properties
presented in this paper are provided for the reader through 
CDS\footnote{Centre de Donn\'ees Astrophysiques de Strasbourg, {\tt 
http://cdsarc.u-strasbg.fr/CDS.html}}, or directly from the authors.
The parameter $\eta$ is a linear combination of the first two 
projections derived from the Principal Component Analysis of the 2dFGRS 
spectra. This parameter is found to be a measure of galaxy spectral 
type, i.e., a measure of the average emission/absorption line strength
of a galaxy (see Madgwick et al. 2002 for a detailed discussion). 
As one may expect, the final emission line galaxy sample contains 
galaxies with bluer colours and later spectral types than the bulk 
of galaxies in the parent 2dFGRS sample. As a result of the selection 
procedure, the redshifts in the final emission line galaxy sample do 
not exceed $z\sim 0.13$, with a median around $z\sim 0.06$.
The observed gaps in the redshift distribution occur when one or more 
of the emission lines lie close to a night-sky emission line, reducing 
the signal-to-noise ratio around that line, and hence the quality of 
the line fitting, which cause them to be excluded them from the final 
emission line galaxy sample.

The completeness of the final galaxy sample is difficult to quantify,
and it is possible that the selection procedure could disguise the 
existence of intrinsic correlations between galaxy properties that 
we aim to investigate. 
Therefore it is important to assess to what extent the final sample 
of emission line galaxies covers similar regions in the parameter 
space (e.g., luminosity, colour, spectral type, surface photometry 
properties) as emission line galaxies in the original 2dFGRS 
sample. To ensure that our final sample of emission line galaxies 
is representative of the parent 2dFGRS emission line galaxies, we
compare the distribution of galaxy properties for both samples. 

\begin{figure*}
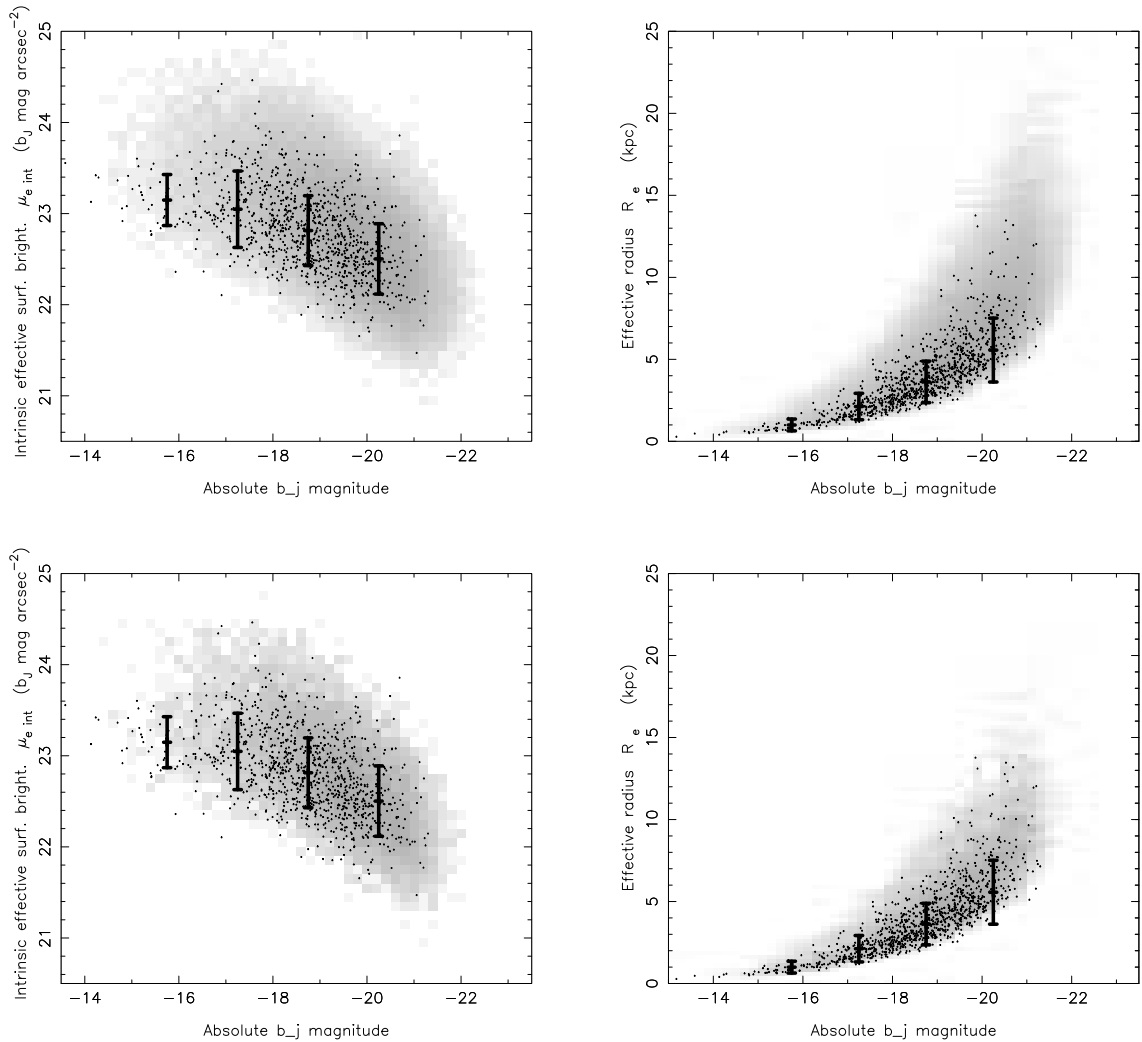

\includegraphics[clip=,angle=270,width=0.45\textwidth]{fig3a.ps}
\includegraphics[clip=,angle=270,width=0.45\textwidth]{fig3b.ps}
\includegraphics[clip=,angle=270,width=0.45\textwidth]{fig3c.ps}
\includegraphics[clip=,angle=270,width=0.45\textwidth]{fig3d.ps}
\caption{Comparisons between the properties of the final 1\,124 emission 
line galaxy sample discussed in the remainder of the paper and those of 
the parent emission line samples. The 2dFGRS emission line galaxy sample 
is shown as a greyscale plot on a logarithmic scale, and the final 
galaxy sample is shown as dots. 
For top panels, galaxies in the parent emission line galaxy sample were 
selected from the 2dFGRS sample requiring $\eta>0$. Top left: intrinsic 
effective surface brightness plotted against absolute $b_j$-band magnitude. 
Top right: physical effective radius against absolute $b_j$-band magnitude. 
For the bottom panel, galaxies in the parent emission line galaxy sample 
were selected are those with $\eta>0$ and ${\rm H\beta}$ equivalent width 
larger than 10\,\AA. Bottom left: similar to the upper left panel. Bottom 
right: similar to the upper right panel. The final 1\,124 galaxy sample 
covers similar ranges of galaxy parameters than the parent sample of 
2dFGRS emission line galaxies with ${\rm H\beta}$ equivalent width larger 
than 10\,\AA.}
\label{comp1}
\end{figure*}

\begin{figure*}
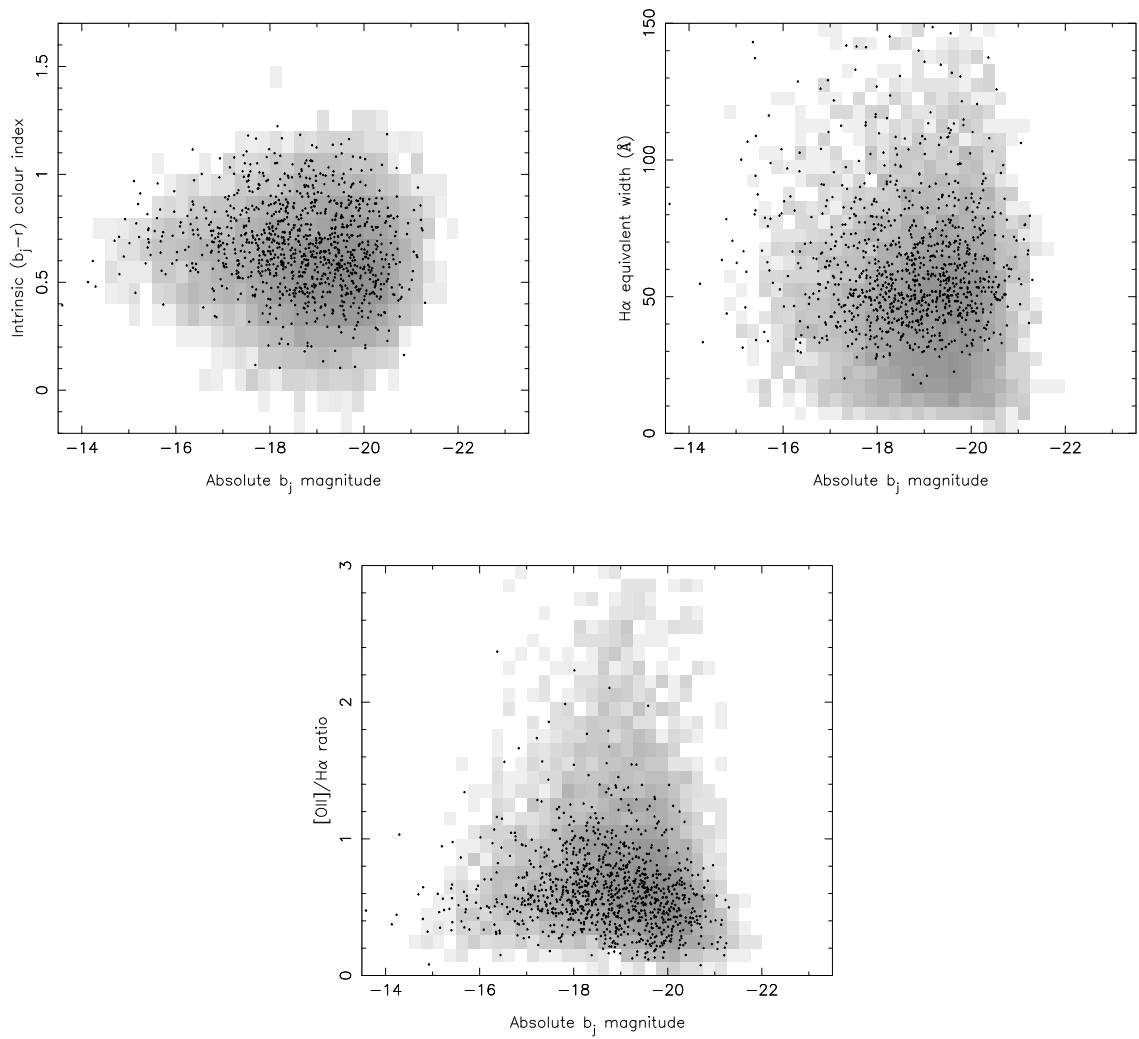

\includegraphics[clip=,angle=270,width=0.45\textwidth]{fig4a.ps}
\includegraphics[clip=,angle=270,width=0.45\textwidth]{fig4b.ps}
\includegraphics[clip=,angle=270,width=0.45\textwidth]{fig4c.ps}
\caption{Comparisons between the variation of ($b_j$-R) colour ($left$), 
${\rm H\alpha}$ equivalent width ($middle$), and observed ${\rm 
[OII]\lambda3727/H\alpha}$ flux ratio ($right$), as a function of $b_j$
magnitude for the final 1\,124 emission line galaxy, shown as dots, and 
the parent emission line sample, shown as a greyscale plot on a logarithmic 
scale. The galaxies in the parent 2dFGRS emission line galaxy sample were 
selected as those with $\eta>0.$ and ${\rm H\beta}$ equivalent width larger 
than 10\,\AA. }
\label{comp2}
\end{figure*}

The parent 2dFGRS emission line galaxy sample was selected as follows.
It contains objects in the final 2dFGRS data release having reliable 
redshifts, for which we have a reliable match between the APM target 
coordinates with the SuperCOSMOS Sky Survey image catalogue (Hambly 
et al. 2001). There is also the additional criterion that heliocentric 
radial velocities ${\rm cz > 1000 km/s}$. This constraint was added 
to guard against including any incorrect velocities caused by having 
a Galactic star superimposed on a galaxy image, or an outright failure 
of the redshift estimation. It also removes very nearby galaxies for 
which the velocity is a poor indicator of distance, and hence guards 
against getting incorrect absolute magnitudes: only a very small 
number of dwarf galaxies with ${M(b_j) > -15}$ mag are rejected (a 
small number because the volume is small, and dwarfs because of the 
apparent magnitude limits of the survey). We have estimated the surface 
photometry parameters for 2dFGRS galaxies using blue images obtained 
from the SuperCOSMOS Sky Survey (see Sect.~\ref{surf_phot} for a 
detailed discussion of surface photometry measurements). The effective 
radius is estimated as the semi-major axis of the ellipse that contains 
half the light of the galaxy, while the effective surface brightness
is the surface brightness at the half-light isophote.

Fig.~\ref{comp1} shows the variation of intrinsic surface brightness and 
physical effective radius as a function of absolute $b_j$-band magnitude 
respectively (see Section~\ref{surf_phot} for details of the calculation 
of these quantities). Our 1\,124 galaxy sample is shown as dots, with 
the bars showing the medians and the standard deviations in 1.5 magnitude 
wide bins. The parent 2dFGRS emission line galaxy sample is shown as a 
greyscale plot on a logarithmic scale. For our sample galaxies, for 
which ${\rm H\alpha}$ equivalent widths are larger than $\sim\,25$\,\AA, 
strong correlations are apparent between surface photometry properties 
and absolute magnitude, i.e., luminous/faint galaxies tend to have on 
average large/small physical sizes and high/low central surface 
brightness. 
The first parent 2dFGRS emission line galaxy sample, shown in the upper 
panels, was constructed from the 2dFGRS sample by selecting galaxies 
with $\eta>0$, i.e., ${\rm H\alpha}$ line in emission, and contains 
71\,207 galaxies. The comparison between the parent 2dFGRS emission 
line galaxy sample, and the final 1\,124 emission line galaxy sample 
shows that $(i)$ the parent emission line galaxy sample includes
luminous galaxies, i.e., brighter than ${\rm M(b_j)\sim\,-21}$, 
that are excluded from the final 1124 emission line galaxy sample, 
and $(ii)$ for a given galaxy magnitude, galaxies with low surface 
brightness and larger effective radii tend to be excluded from our 
final 1124 emission line galaxy sample. This is because galaxies in 
the final sample were selected on the basis of their ${\rm H\beta}$ 
equivalent width being larger than 10\,\AA. To take this selection 
criterion into account, we have constructed a second parent 2dFGRS 
emission line galaxy sample by selecting galaxies from the 2dFGRS 
sample requiring both $\eta > 0$ and ${\rm H\beta}$ equivalent width 
larger than 10\,\AA. This second sample contains 12\,660 galaxies. 
The bottom panels of Fig.~\ref{comp1} show a comparison between our 
1\,124 emission line galaxy sample and the second 2dFGRS parent 
emission line galaxy sample. It shows that emission line galaxies in 
the final emission line galaxy sample cover roughly similar ranges of 
galaxy parameters as the second parent 2dFGRS emission line galaxy 
sample, and that galaxies are distributed similarly in both samples. 
The cut on ${\rm H\beta}$ equivalent width excludes bright/physically 
big galaxies, since galaxies with low emission line equivalent width 
tend to dominate the bright end of the galaxy luminosity function 
(e.g., Salzer et al. 1989; Kong et al. 2002). Despite this cut, 
the final 1\,124 emission line sample covers a large range of galaxy 
luminosities, i.e., 7 magnitudes, similar to the magnitude range 
covered by NFGS sample (Jansen et al. 2000) or 15R-North galaxy 
redshift survey sample (Carter et al. 2001). This is attributed to 
the large scatter that affects the galaxy luminosity versus emission 
line equivalent width relation (e.g., Jansen et al. 2000). 
The distributions of the final 1\,124 emission line galaxy sample 
and the second parent 2dFGRS emission line galaxy sample are similar 
in both ($b_j$-R) vs. ${\rm \mu(b_j)}$ and ($b_j$-R) vs. ${\rm R_e}$
diagrams.

Fig.~\ref{comp2} shows the variation of ($b_j$-R) colour, 
${\rm H\alpha}$ equivalent width, and extinction-uncorrected 
${\rm [OII]\lambda3727/H\alpha}$ flux ratio (see below for a 
detailed discussion of the procedure we use to estimate this ratio) 
as a function of $b_j$-band magnitude for the final 1\,124 galaxy 
sample and the parent 2dFGRS emission line galaxy sample. Galaxies 
in the parent 2dFGRS sample were selected by having $\eta>0$ and 
${\rm H\beta}$ equivalent width larger than 10\,\AA. Again, the 
final 1\,124 galaxy sample is distributed similarly to the 2dFGRS 
parent sample of emission line galaxies with ${\rm H\beta}$ equivalent 
width larger than 10\,\AA. The parent sample of emission line galaxies 
with ${\rm H\beta}$ equivalent width larger than 10\,\AA\ contains 
galaxies with low ${\rm H\alpha}$ equivalent width, and high observed 
${\rm [OII]\lambda3727/H\alpha}$ flux ratio that are not present in 
the final 1\,124 emission line galaxy sample. However, these galaxies
represent less than 1\% of the parent sample.
The similar distributions of both emission line galaxy samples suggest 
that our final 1\,124 galaxy sample is representative of the parent 
sample of 2dFGRS emission line galaxies with ${\rm H\beta}$ equivalent 
width larger than 10\,\AA\ in terms of its luminosity, colour, and 
surface photometry properties. Also, the 2dFGRS spectroscopic sample 
is representative of the complete 2dFGRS magnitude-limited photometric 
sample down to the surface brightness limit of the 2dFGRS, i.e., ${\rm 
\sim\,24.5 \, b_j\,mag\,arcsec^{-2}}$ (Cross et al. 2001). Hence, the 
final 1\,124 emission line galaxy sample is fairly representative of 
the vigorously star forming, i.e., ${\rm H\beta}$ equivalent width 
larger than 10\,\AA, local emission line galaxies, within the 2dFGRS 
limits. The correlations we aim to investigate are expected then not 
to be severely biased by the selection procedure of the final sample 
of 1\,124 emission line galaxies. However, we can not rule out the 
possibility that we are missing faint galaxies with strong emission 
lines, but with low surface brightness, i.e., close to or lower than 
the 2dFGRS surface brightness limit.

\section{Aperture effects}
\label{aper}

The fibres in the 2dF instrument cover a 2.1 arcsecond diameter region 
of each galaxy, and this small aperture coverage of galaxies may bias 
the distribution of galaxy properties, and the estimate of the emission 
line properties (e.g., Kochanek et al. 2001). The aperture in fibre-fed 
spectroscopy is usually centred on the inner part of the galaxies so 
that the nuclear light is collected. For 2dFGRS, the internal precision 
with which the fibres are aligned with the galaxy centre is ${\rm 
0.16\,arcsec}$ on average, with no fibres outside ${\rm 0.3\,arcsec}$ 
(Colless et al. 2001). 
The absolute accuracy of the input astrometry is ${\rm \sim\,0.5\,arcsec}$ 
(Maddox et al. 1990a). The fraction of the light from outer parts of 
a galaxy will depend on its redshift, intrinsic size, and surface
brightness profile, as well as the size of the fibre and seeing during 
the observation.  The bias introduced by such observational procedure
depends also on the morphological type of the observed galaxies. The
larger the bulge-to-disk ratio, the more serious may be the potential
bias. Even though the distribution of star-forming regions tend to be
centrally distributed in bulge-dominated galaxies, luminous star
forming regions tend to be located in the outer regions of the
disk. The major caveat might come from irregular galaxies where star
forming regions can be found anywhere.  Aperture effects are therefore
a concern, and we must assess how close the spectra of our sample
galaxies are to fully integrated galaxy spectra. We first consider the
surface brightness profiles of the galaxies, to estimate the fraction
of a galaxy's light that is sampled by the fibre. We then search for
any correlation between emission line properties and the fraction of
light sampled.

\subsection{Surface photometry of 2dFGRS galaxies} 
\label{surf_phot}

Surface photometry was carried out using blue image data from the
United Kingdom Schmidt Telescope obtained from the SuperCOSMOS Sky
Survey (Hambly et al. 2001, Hambly, Irwin \& McGillivray 2001). 
Data were downloaded from the public Survey server in 
Edinburgh\footnote{http://www-wfau.roe.ac.uk/sss/} for a 2.0 arcmin 
square region around each galaxy.

The SExtractor program (Bertin 1998, Bertin \& Arnouts 1996) was run 
over each data file and the SExtractor object corresponding to the 
2dF target was identified, on the basis of a close match to the 
celestial coordinates of the 2dFGRS target. For each 2dF target, 
SExtractor provided the object centroid coordinates, an overall image 
ellipticity and the orientation of the object. This analysis used only 
those pixels having a surface brightness brighter than 25.0 $b_j$ mag 
arcsec$^{-2}$. A series of concentric elliptical annuli were then 
defined on a linear scale for each 2dFGRS target which were centred on 
the image centroid of the target and had the overall image ellipticity 
and orientation. The mean SuperCOSMOS intensity was then measured 
within each annulus. The results from these annuli provided a surface 
brightness profile in the form of the mean intensity as a function of 
the semi-major axis of the annuli. 

Exponential surface brightness profiles were fitted to the surface 
brightness profiles weighting each data point according to an estimate 
of the error in the intensity. These fits were performed between 
intensities corresponding to 22.3 and 27.0 $b_j$ mag arcsec$^{-2}$. 
The bright limit was imposed to avoid problems associated with 
photographic saturation in the UKST survey data (e.g Maddox et al. 
1990ab, Shao et al. 2003). Data points within 1.5~arcsec of the image 
centre were excluded to avoid any problems caused by the presence of 
a nucleus or by seeing. The fitted profile is characterised by the 
central surface brightness and the exponential scale length along the 
major axis. As the method avoids the saturated higher surface brightness 
regions of galaxy images, in the case of spiral galaxies it is mostly 
sensitive to their discs. The central surface brightness is therefore 
an extrapolation to the image centre of the fitted profile. 

The SuperCOSMOS intensities were converted to magnitude surface
brightnesses using calibrations for each UKST plate derived from the
public SuperCOSMOS Sky Survey total magnitudes. The photometric zero
point of the SuperCOSMOS data for each UKST plate was selected to
minimise the differences between the total magnitude under the fitted
exponential profile and the SuperCOSMOS Sky Survey total $b_j$
magnitudes for all 2dFGRS targets within that UKST field. No attempt
was made to account for photometric variations across individual UKST
plates: the SuperCOSMOS Sky Survey pixel data have already been corrected
for vignetting effects, and no strong evidence of residual effects were
found during a comparison of the fitted profile data with total
magnitudes from the APM Catalogue (Maddox et al. 1990a). The magnitude
surface brightnesses of the fitted exponential profiles have therefore
been put on to the SuperCOSMOS Sky Survey scale. The root-mean-square 
difference between the total magnitude under the fitted exponential 
profiles and the total magnitudes from the SuperCOSMOS Sky Survey 
Catalogue is 0.15~mag, which will be caused by factors including a 
failure to account for bulges of spiral galaxies, the $r^{1/4}$ profiles 
of ellipticals and peculiar galaxy morphologies. This figure is slightly 
larger for lower redshifts and for lower surface brightnesses, but does 
not vary with colour. The data of Hambly, Irwin \& MacGillivray (2001) 
indicate that the absolute calibration of the SuperCOSMOS Sky Survey 
Catalogue blue magnitudes is accurate at the 0.1--0.2~mag level. 

The surface photometry gave the inferred central surface brightness 
$\mu_{\circ,\:\rm obs}$ and the angular scale length. To correct for 
cosmological surface brightness dimming, the intrinsic central 
surface brightness was computed as: 
\begin{equation}
  \mu_{\circ,\:\rm int} \; = \; \mu_{\circ,\:\rm obs} - 
                          2.5 \log_{10}(1+z)^4 - k(z) - A_{\rm Gal}
\end{equation}
where $z$ is the redshift, $k(z)$ is the $k$-correction and $A_{Gal}$ 
is the interstellar Galactic foreground extinction. The $k$-correction 
was computed for the observed 2dF redshift using the Poggianti (1997) 
results for the $b_j$ photographic band, determining the galaxy type 
from the Madgwick et al. (2002) $\eta$ parameter. The physical scale 
length was obtained from the angular scale length using the angular 
diameter distance computed from the redshift. The mean value of the 
$b_j$ $k$-correction for the emission-line galaxies was 0.15~mag. 
No corrections were made for inclination and internal reddening.

\begin{figure}
\includegraphics[clip=,width=0.45\textwidth]{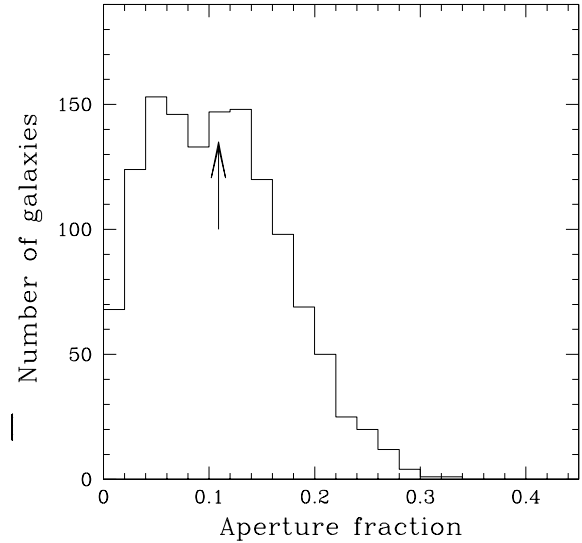}
\caption{Distribution of the fraction of the total galaxy light 
observed by the fibre for the galaxy sample discussed in the paper. 
The vertical arrow indicates the location of the mean of the covering 
fraction distribution, i.e., $11\%$. }
\label{dist_ap}
\end{figure}

\begin{figure*}
\includegraphics[clip=,width=0.45\textwidth]{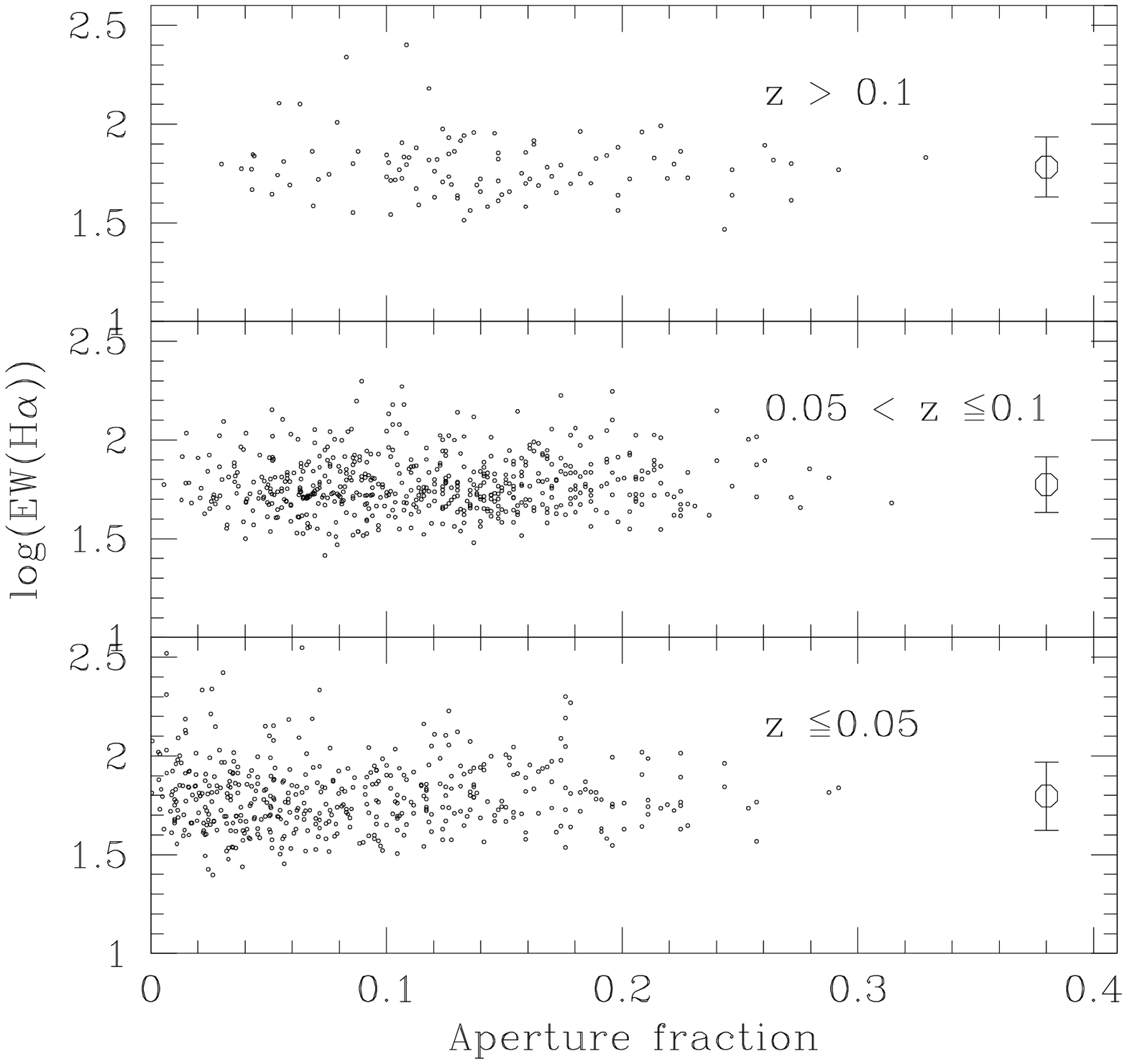}
\includegraphics[clip=,width=0.45\textwidth]{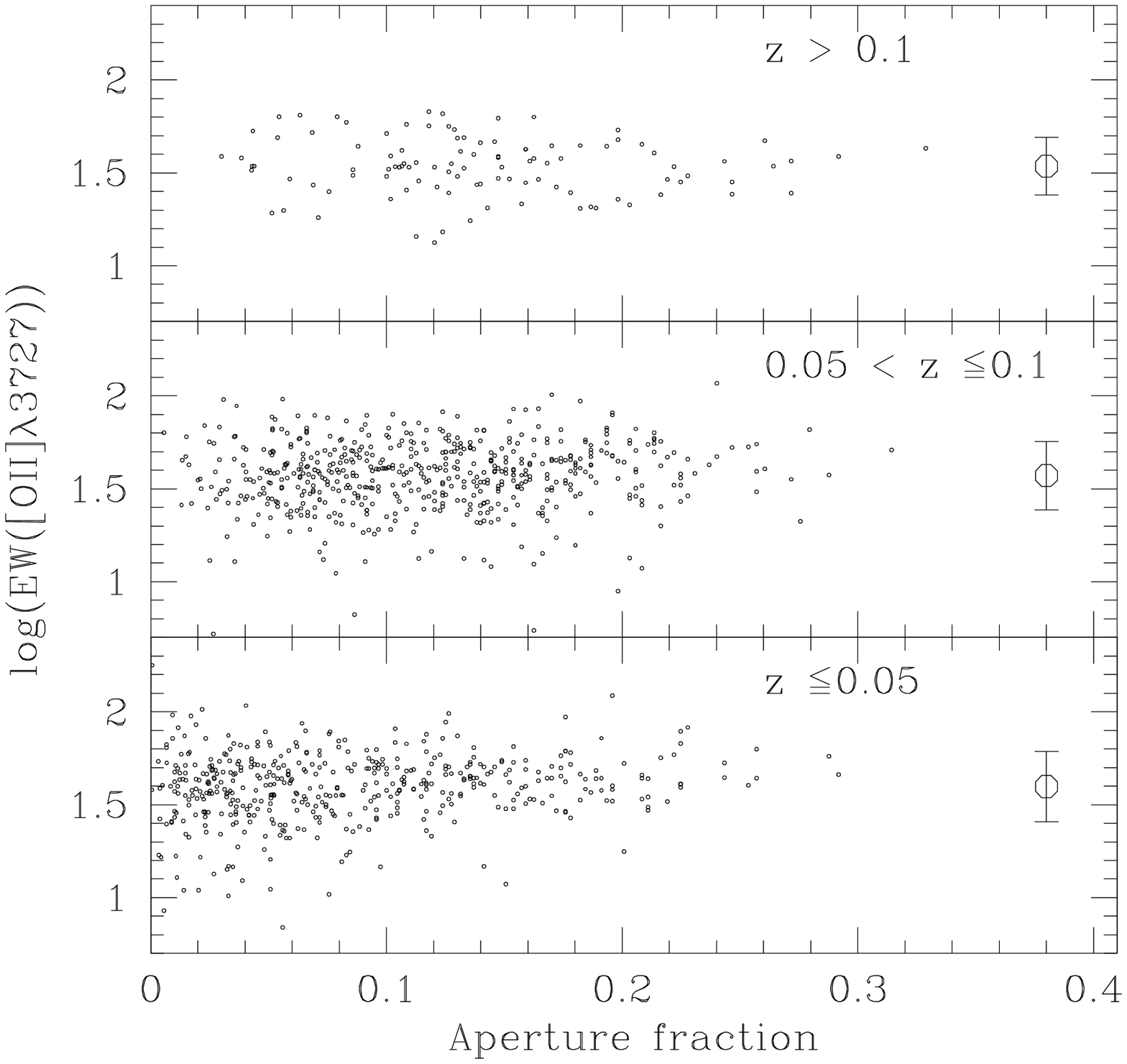}
\includegraphics[clip=,width=0.45\textwidth]{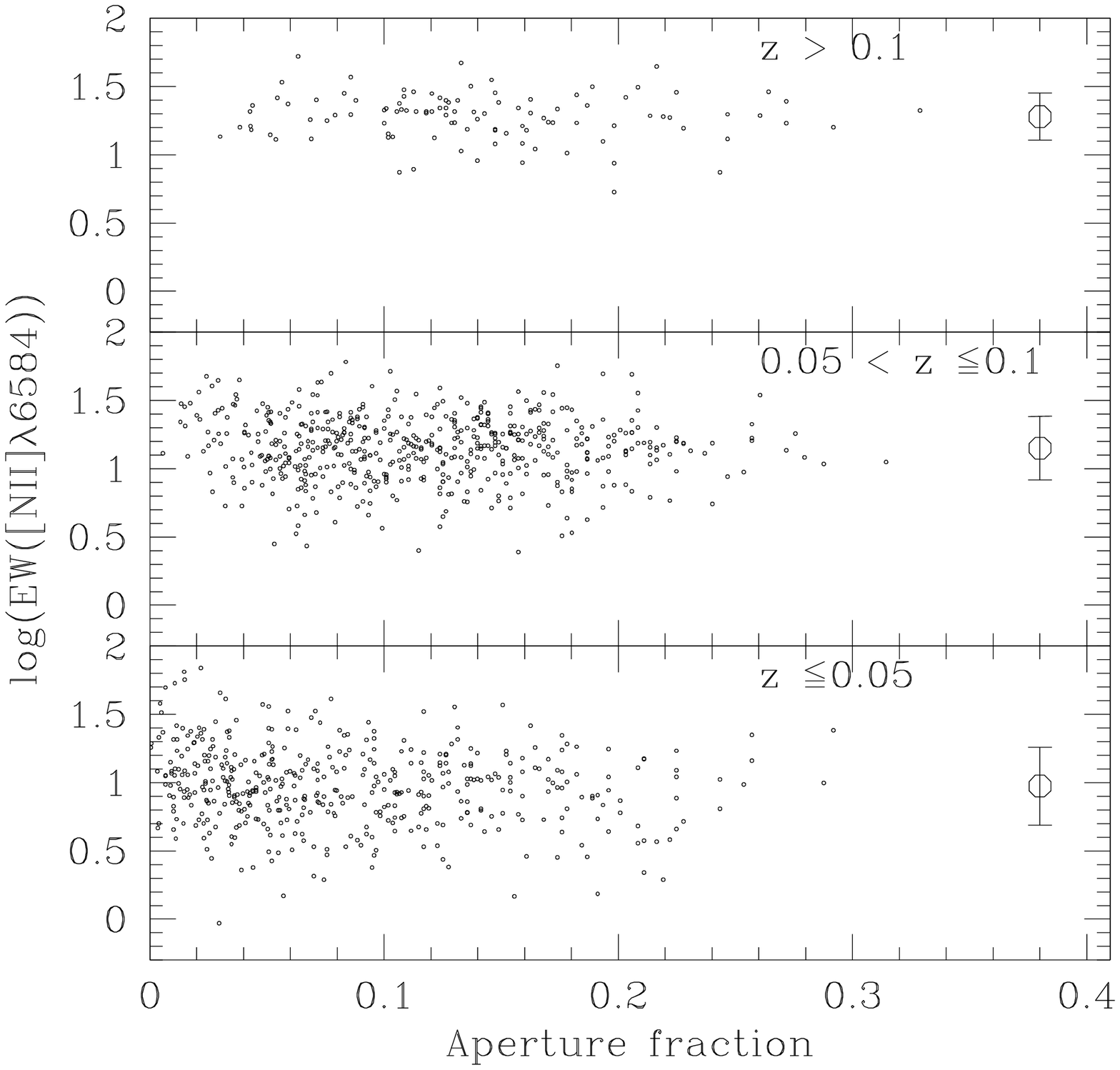}
\includegraphics[clip=,width=0.45\textwidth]{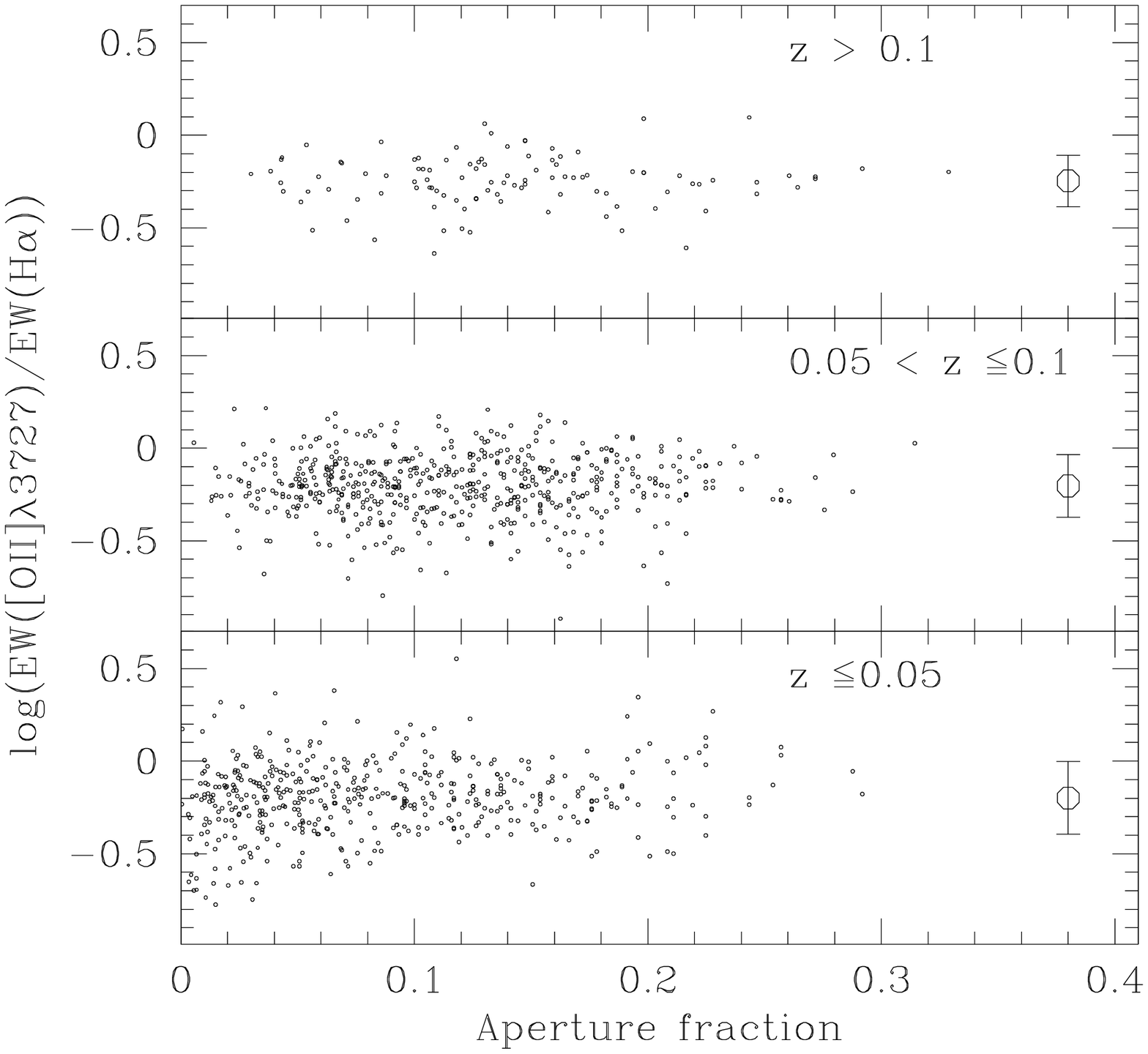}
\caption{A test for aperture effect. Different panels show the 
relationships between the equivalent widths of ${\rm H\alpha}$ (top 
left), ${\rm [OII]\lambda3727}$ (top right), ${\rm [NII]\lambda6584}$ 
(bottom left), and the ratio of ${\rm [OII]\lambda3727}$ and ${\rm 
H\alpha}$ equivalent widths (bottom right) and the aperture fraction 
for the galaxy sample discussed in the paper for different redshift bins. 
The large circle and the error bar in each panel show the median and 
the standard deviation of the distribution of emission line equivalent 
widths and the equivalent width ratio in each redshift bin.}
\label{ap_ew}
\end{figure*}

\subsection{Aperture coverage}
\label{aper2}

We use the surface photometry, discussed in Sec.~\ref{surf_phot} to
estimate the fibre covering fraction for the emission line galaxies 
by measuring the ratio of the light within the fibre aperture, 
centred on the nucleus, to the total luminosity. Fig.~\ref{dist_ap} 
shows the distribution of the covering fraction for the galaxy sample. 
It shows that the average spectrum in our sample contains $11\%$ of 
the total flux of the galaxy, with a standard deviation of $6\%$; 
the median of the covering fraction distribution is 10\%. The average 
fraction of galaxy light collected by the fibres depends on redshift: 
the 14\% aperture covering fraction for $z>0.1$ galaxies is a factor two 
larger than for $z<0.05$ galaxies. Intrinsically brighter galaxies tend 
to be physically larger than fainter galaxies, but for magnitude-limited 
survey such as the 2dFGRS they also tend to be found out to larger 
distances, where fainter galaxies are lost from the sample, so the 
projected fibre aperture is larger. Thus for a magnitude-limited sample, 
the fraction of galaxy light seen by the fibres is not a strong function 
of absolute magnitude (see Fig.\,9 of Tremonti et al. 2004 for a similar 
conclusion using a sample of emission line galaxies drawn from the SDSS).

To make sure that the aperture size is not biasing the properties of
our galaxy sample, we examine the galaxy properties as a function of
the fraction of galaxy light collected through the fibres. 
Fig.~\ref{ap_ew} shows the relationship between the aperture fraction 
and the equivalent width of ${\rm [NII]\lambda6584}$, ${\rm 
[OII]\lambda3727}$, ${\rm H\alpha}$, and ${\rm [OII]\lambda3727/H\alpha}$ 
equivalent width ratio, for different redshift bins. The large open 
circles and the bars show the medians and the standard deviations of 
the emission line equivalent width distributions and equivalent width 
ratio distribution for each redshift bin. The figure shows that the 
equivalent widths do not show any sizeable dependence on the fibre 
covering fraction, independently of redshift. The emission line 
equivalent widths and equivalent width ratio show no trend with 
redshift. The distributions of emission line equivalent widths for 
galaxies with redshifts $z\ge 0.05$, a redshift limit recommended by 
Zaritsky et al. (1995) to minimize the effects of the aperture bias, 
do not show any trend with the observed fraction of the galaxy light. 
Consequently, there is no evidence that the properties of our normal
emission line galaxy sample are systematically biased. We conclude 
that the galaxy sample may be used with confidence to study the 
properties of emission line galaxies.

\section{The $\rm [OII]\lambda3727/{\rm H\alpha}$ flux ratio}
\label{oii_ha}

In this section, we will investigate the sensitivity of ${\rm 
[OII]\lambda3727/H\alpha}$ emission line flux ratio to galaxy and 
interstellar emitting gas properties.

\subsection{Flux calibration and reddening } 
\label{o2ha_ebmv}

The relative flux calibration, over the whole spectral coverage of 
the 2dF spectrograph is uncertain; thus the flux ratio of two distant 
lines, such as ${\rm [OII]\lambda3727}$ and ${\rm H\alpha}$, may not 
be accurately estimated and may also be subject to systematic errors. 
Fortunately, using equivalent widths and broad-band photometry, one 
can accurately estimate the flux ratio.
Let us write the emission line equivalent width as a ratio between 
the emission line flux and the adjacent continuum flux in the observed
spectrum, so the extinction-corrected ${\rm [OII]\lambda3727/H\alpha}$
emission line flux ratio is given by:

\begin{eqnarray}
I([OII])/I(H\alpha) = EW([OII])/EW(H\alpha) 
                      \times F_{c,[OII]}/F_{c,H\alpha} \nonumber \\
            \times 10^{0.4\times E(B-V)(\kappa(OII)-\kappa(H\alpha))}
\end{eqnarray}

where $F_{c,[OII]}$, and $F_{c,H\alpha}$ are the continuum flux 
adjacent to ${\rm [OII]\lambda3727}$ and ${\rm H\alpha}$ respectively,
$E(B-V)$ is the colour excess. We approximate the ratio between the 
continuum flux by a colour term: ${\rm \log(F_{c,[OII]}/F_{c,H\alpha})
\,=\,0.4[z_p-(b_j-R)}]$, where $z_p=0.6$.

The colour excess, $E(B-V)$, from obscuration by dust can be estimated 
from the observed ratio of ${\rm H\alpha}$ and ${\rm H\beta}$ line 
fluxes for each galaxy using the relation:
\begin{equation}
E(B-V)\,=\,1.086\tau_{V}/R_{V} 
\end{equation}
where
\begin{equation}
\tau_{V}\,=\,\frac{\ln[F(H\alpha)/F(H\beta)]-\ln[I(H\alpha)/I(H\beta)]}
                  {\kappa_{H\alpha}-\kappa_{H\beta}}
\end{equation}
${\rm I(H\alpha)/I(H\beta)}$ is the intrinsic Balmer line flux ratio,
${\rm F(H\alpha)/F(H\beta)}$ is the observed Balmer line flux ratio, 
$\tau_{V}$ is the effective $V$-band optical depth, and $\kappa_{\lambda}$  
is the optical interstellar extinction curve. We adopt the Milky Way 
interstellar extinction law of Cardelli, Clayton, \& Mathis (1989), 
with $R_V=3.1$.
We make the stellar absorption correction to ${\rm H\alpha/H\beta}$ 
flux ratio on a galaxy-by-galaxy basis by fitting Gaussian profiles to 
both an absorption and emission component for ${\rm H\beta}$. We also 
correct the ${\rm H\alpha/H\beta}$ flux ratio for Galactic extinction 
using values taken from Schlegel, Finkbeiner, \& Davis (1998) extinction 
maps.
We assume an intrinsic ratio of ${\rm I(H\alpha)/I(H\beta)=2.85}$, 
corresponding to the case B recombination with a temperature of 
${\rm T=10^4 K}$, and a density of ${\rm n_e\sim 10^2-10^4\, cm^{-3}}$
(Osterbrock 1989). The different extinction laws available in the 
literature show similar behaviour in the optical, making the results 
of our subsequent analysis independent of the chosen extinction law.

\begin{figure}
\includegraphics[clip=,width=0.45\textwidth]{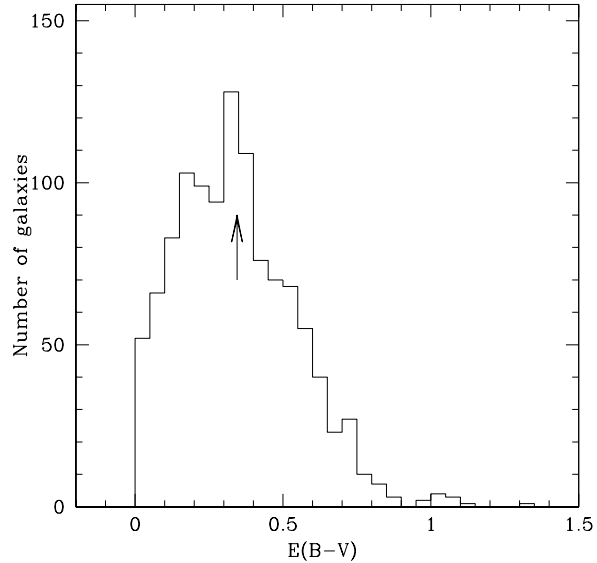}
\caption{Distribution of the colour excess $E(B-V)$ for the galaxy sample.
The vertical arrow shows the location of the mean colour excess for the
sample, i.e., $E(B-V)=0.34\pm 0.01$. }
\label{ext_dist}
\end{figure}

\begin{figure}
\includegraphics[clip=,width=0.45\textwidth]{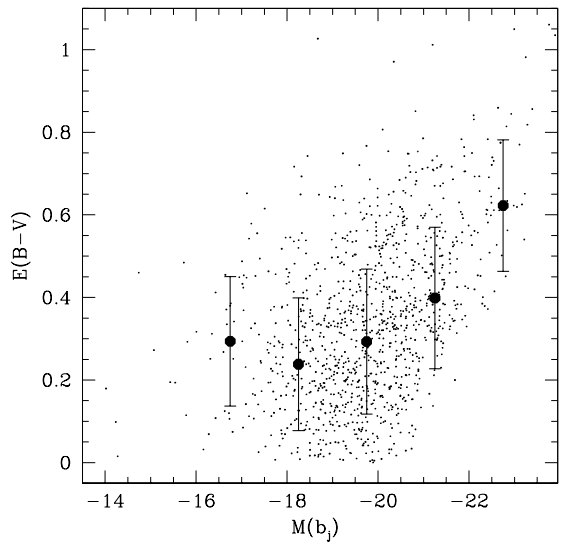}
\caption{Relationship between Balmer-decrement derived colour excess 
and the absolute magnitude M($b_j$). Large filled circles and bars show
the mean and the standard deviation of the colour excess distributions 
in 1.5 magnitude wide bins.}
\label{ext_mag}
\end{figure}

Fig.~\ref{ext_dist} shows the distribution of inferred colour excess
for our galaxy sample. The mean $E(B-V)$ for the galaxies in our
sample, after correcting for Galactic extinction, is $0.34\pm 0.01$,
and a median of $0.33\pm 0.01$, with a standard deviation of $0.2$ 
magnitude. The mean value found for the galaxy sample is consistent 
with the widely used average colour excess, $E(B-V)\sim 0.3$, for 
${\rm H\alpha}$ measurements of star forming galaxies (e.g., 
Nakamura et al. 2003, Hopkins et al. 2003, Kewley et al. 2004).

Fig.~\ref{ext_mag} shows the relationship between the colour 
excess and absolute $b_j$-band magnitude. Large filled circles 
and associated bars show the mean and the standard deviation 
of the colour excess distributions in 1.5 magnitude 
wide bins. The Spearman correlation coefficient is $-0.45$. 
The two-sided probability of obtaining this value by chance is 
almost zero. This indicates that the colour excess and the 
absolute galaxy magnitude are correlated; i.e., bright galaxies 
tend to be more affected by internal extinction than faint 
galaxies, and that there is a large scatter about this trend.
The absence of galaxies in the lower right corner of the plot, 
i.e., the lack of bright galaxies with low colour excess, is 
unlikely to be due to a selection effect as the selection of 
sample galaxies was based uniquely on the strength of emission 
lines compared to the continuum, not on galaxy absolute magnitude 
and/or emission line ratios. Samples of emission line galaxies 
selected with different ${\rm H\beta}$ equivalent width cuts 
between 10\,\AA\ and 20\,\AA\ do not show a larger zone of 
exclusion in the lower right corner of Fig.~\ref{ext_mag}.

If the dust were smoothly distributed throughout the galaxies, light
from the general stellar population would be obscured, and we would
expect to see a lower central surface brightness in galaxies with a
larger colour excess.  We find no significant correlation between 
the colour excess and the galaxy central surface brightness or the
physical effective radius. This suggests that the obscuring dust in 
galaxies is not distributed in the same way as the stellar light,  
it is concentrated close to the sources of line emission (e.g., 
Stasi\'nska \& Sodr\'e 2001), and/or it has different scale-height 
to stars. 

\begin{figure}
\includegraphics[clip=,width=0.45\textwidth]{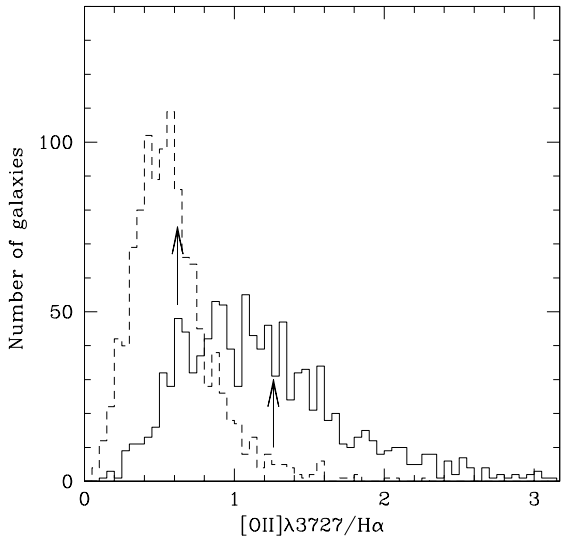}
\caption{Distribution of the ${\rm [OII]\lambda3727/H\alpha}$ flux ratio. 
The continuous line shows the distribution of the extinction-corrected 
ratio, while the dashed line shows the ratio before extinction correction. 
The vertical arrows show the means of the extinction-uncorrected and 
extinction-corrected ratio distributions, i.e., $0.62\pm 0.02$ and 
$1.26\pm 0.03$ respectively. }
\label{o2ha_distr}
\end{figure}

\begin{figure}
\includegraphics[clip=,width=0.45\textwidth]{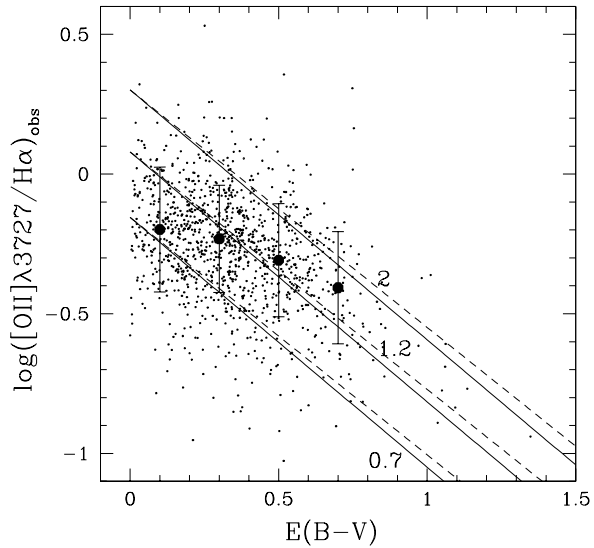}
\caption{The relationship between the colour excess E(B-V) derived from 
the Balmer decrement and the observed ${\rm [OII]\lambda3727/H\alpha}$ 
ratio. Large circles and bars show the means and the standard deviations
of the observed ${\rm [OII]\lambda3727/H\alpha}$ ratio in 0.2 magnitude
wide bins. The continuous lines indicate the trend expected from the 
obscuration curve of Cardelli et al. (1989) for intrinsic ${\rm 
[OII]\lambda3727/H\alpha}$ flux ratios of 0.7, 1.2, and 2.0 from bottom
to top, respectively. The dashed lines show the similar trend but using 
the obscuration curve of Seaton (1979).}
\label{o2ha_redd}
\end{figure}

Fig.~\ref{o2ha_distr} shows the distribution of the observed and the
extinction-corrected ${\rm [OII]\lambda3727/H\alpha}$ emission line
flux ratio. The mean value of the extinction-corrected ratio is 
$1.26\pm 0.02$, compared to $0.62\pm 0.02$ of the observed emission
line ratio. Both observed and extinction-corrected mean ratios for
our emission line galaxy sample are comparable to the same mean 
ratios for NFGS galaxies (Kewley et al. 2004); this is reasonable 
since both galaxy samples select roughly similar emission line 
galaxies.
However the observed ratio for our sample is different from the 
values seen in the UCM galaxy sample (Arag\'on-Salamanca et al. 2004), 
and radio-detected galaxies in the First Data Release of the SDSS 
(Hopkins et al. 2003). On the other hand, the mean value of the 
extinction-corrected ratio is comparable to what is derived for 
the UCM samples. Note that the Hopkins et al. (2003) sample has the 
lowest observed mean ${\rm [OII]\lambda3727/H\alpha}$ ratio; this 
is understandable since radio-selected samples tend to be less biased 
against galaxies with a larger dust content than optically- or 
${\rm H\alpha}$-selected galaxy samples. This confirms the Jansen et al.
(2001) finding that an important factor leading to different emission 
line ratios in different galaxy samples is the sample-dependent mean 
dust extinction. Thus using ${\rm [OII]\lambda3727}$ as a star 
formation rate indicator requires calibration in a reddening-independent 
way (see also Kewley et al. 2004).

Fig.~\ref{o2ha_redd} shows the relationship between internal dust 
reddening, in terms of colour excess, E(B-V), and the logarithm of 
the observed ${\rm [OII]\lambda3727/H\alpha}$ ratio. Large circles 
and bars show the means and the standard deviations of the observed 
${\rm [OII]\lambda3727/H\alpha}$ ratio in 0.2 magnitude wide bins.
The trend indicated by large solid circles does not change if the
medians are used instead of the means. 
The Spearman correlation coefficient is $-0.35$, with the two-sided 
probability of obtaining this value by chance almost zero, i.e., 
$\sim 8\times\,10^{-32}$. This indicates a statistically significant 
correlation between the colour excess, E(B-V), and emission line 
${\rm [OII]\lambda3727/H\alpha}$ flux ratio, consistent with what
was found for NFGS galaxies (Jansen et al. 2001; Kewley et al. 2004). 
We found no convincing relationship between galaxy luminosity and 
extinction-corrected ${\rm [OII]\lambda3727/H\alpha}$ ratio for our 
galaxy sample, similar to what is observed for UCM galaxies 
(Arag\'on-Salamanca et al. 2005, see also Hopkins et al. 2003). 
However, Jansen et al. (2001) found that after correcting for the 
internal extinction, a weak correlation still exists between the 
${\rm [OII]\lambda3727/H\alpha}$ ratio and the galaxy luminosity. 
They interpret this correlation as an indication of the sensitivity 
of emission line ratio to gas-phase abundance. 
If a common extinction law is valid for all the galaxies in the 
sample, there should be a simple relationship between the observed 
${\rm [OII]\lambda3727/H\alpha}$ flux ratio and the colour excess.
The solid lines in Fig.~\ref{o2ha_redd} show the relationships 
expected using the extinction law of Cardelli et al. (1989) for 
different values of the intrinsic ${\rm [OII]\lambda3727/H\alpha}$ 
flux ratio. The dashed lines show the expected relationships if 
the extinction law of Seaton (1979) is used. The fact that the 
predicted relationship using the Seaton (1979) extinction law
is not significantly steeper than what is expected using the
extinction law of Cardelli et al. (1989) suggests that the 
presence or absence of a correlation between absolute magnitude 
and the extinction-corrected ${\rm [OII]\lambda3727/H\alpha}$ 
is not tied strongly to the adopted extinction law. It is possible
that the large scatter of 2dFGRS data may mask a correlation.

\subsection{ ${\rm [OII]\lambda3727/H\alpha}$ ratio and metal 
             abundance}
\label{o2ha_oh_sec}
 
To what extent does the systematic variation of galaxy chemical 
abundance regulate the variation of the ${\rm 
[OII]\lambda3727/H\alpha}$ ratio?
Because of the sensitivity of ${\rm [OII]\lambda3727}$ to metallicity, 
one may expect that the ${\rm [OII]\lambda3727/H\alpha}$ ratio may be 
related to the metal-content of the star-forming region. Unfortunately, 
2dFGRS spectra do not have the required S/N to accurately measure 
the needed emission lines to estimate electronic temperature-based 
abundances.  Without a reliable electron temperature diagnostic, we 
have estimated the gas-phase oxygen abundance using the so-called 
strong emission line method first proposed by Pagel et al. (1979), 
and extensively used in the literature (e.g., Dopita \& Evans 1986, 
Zaritsky et al. 1994, Contini et al. 2002, Melbourne \& Salzer 2002, 
Pettini et al. 2001, Kobulnicky et al. 2003). 
This approach is based on the idea that strong lines, i.e., ${\rm 
[OII]\lambda3727}$, ${\rm [OIII]\lambda4959,\lambda5007}$, and ${\rm 
H\beta}$, contain enough information to get an accurate estimate of 
the oxygen abundance (McGaugh 1991). This is done through the so-called 
parameter $R_{23}$, introduced by Pagel et al. (1979), and defined as: 
${\rm R_{23}= ([OIII]\lambda4959,\lambda5007+[OII]\lambda3727)/H\beta}$. 
The ${\rm R_{23}}$ parameter is estimated usually from emission line 
flux ratio. Recently, Kobulnicky \& Phillips (2003) have shown that 
the use of equivalent widths instead of fluxes to derive $R_{23}$ 
gives similar results. Due to the limited quality of the relative flux 
calibration over the whole spectral range covered by the 2dF spectra, 
we prefer to use the equivalent widths to estimate $R_{23}$ rather than 
emission line fluxes. 

\begin{figure}
\includegraphics[clip=,width=0.45\textwidth]{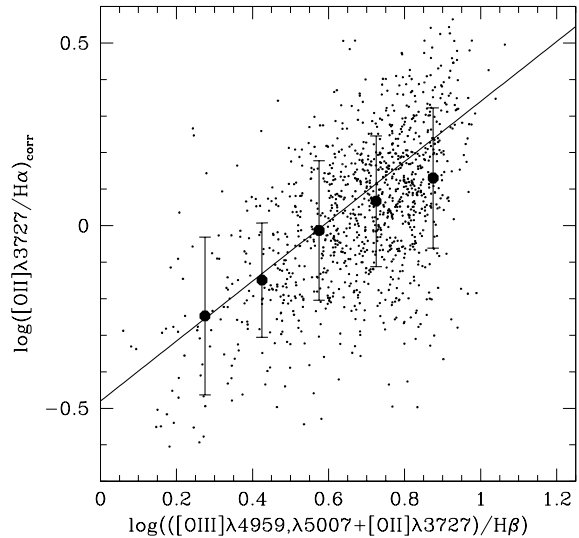}
\caption{The relationship between the reddening-corrected ${\rm 
[OII]\lambda3727/H\alpha}$ ratio and the abundance sensitive $R_{23}$ 
parameter. Large filled circles and bars show the means and the standard 
deviations of logarithmic reddening-corrected ${\rm 
[OII]\lambda3727/H\alpha}$ ratio distributions in 0.15 dex wide bins. 
The solid line is a linear fit from Jansen et al. (2001). }
\label{o2ha_r23}
\end{figure}

\begin{figure}
\includegraphics[clip=,width=0.45\textwidth]{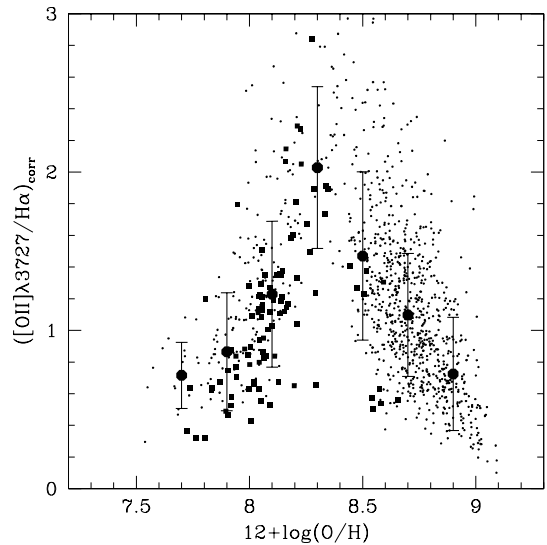}
\caption{The relationship between the reddening-corrected ${\rm 
[OII]\lambda3727/H\alpha}$ ratio and the oxygen abundance in units of 
${\rm 12+\log(O/H)}$. The abundances are calculated using the McGaugh 
(1991) calibration of the $R_{23}$ method (see text for more details). 
Filled circles and associated error bars show the means and the standard
deviations of ${\rm [OII]\lambda3727/H\alpha}$ ratio distributions as a 
function of oxygen abundance in 0.2 dex wide bins. Filled squares show 
galaxies with high ionization-sensitive ratios, i.e., ${\rm 
\log([OIII]\lambda5007/H\beta)\ge\,0.5}$. These galaxies tend to have, 
at a given oxygen abundance, lower reddening-corrected ${\rm 
[OII]\lambda3727/H\alpha}$ ratios than galaxies with low and intermediate 
${\rm \log([OIII]\lambda5007/H\beta)}$ ratios (see text for more details). }
\label{o2ha_oh}
\end{figure}

\begin{figure}
\includegraphics[clip=,width=0.45\textwidth]{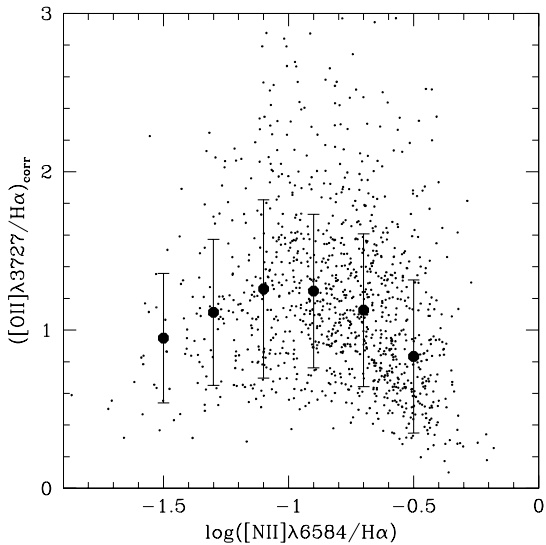}
\caption{The relationship between the reddening-corrected 
${\rm [OII]\lambda3727/H\alpha}$ ratio and the abundance sensitive 
emission line ratio ${\rm [NII]\lambda6584/H\alpha}$. Filled circles 
and associated error bars show the means and the standard deviations 
of ${\rm [OII]\lambda3727/H\alpha}$ ratio distributions in 0.2 dex 
wide bins.}
\label{o2ha_n2ha}
\end{figure}

Fig.\,\ref{o2ha_r23} shows the relationship between the intrinsic
${\rm [OII]\lambda3727/H\alpha}$ flux ratio and the abundance-sensitive 
$R_{23}$ parameter. Large filled circles and bars show the means and 
the standard deviations of logarithmic reddening-corrected ${\rm 
[OII]\lambda3727/H\alpha}$ ratio distributions in 0.15 dex wide bins.
The solid line is the linear fit to the NFGS galaxy sample (Jansen et 
al. 2001). The Spearman correlation coefficient is $-0.55$, with the 
two-sided probability of obtaining this value by chance being almost 
zero. This indicates that intrinsic ${\rm [OII]\lambda3727/H\alpha}$ 
flux ratio is correlated with the abundance-sensitive $R_{23}$ parameter 
with a large statistical significance. The observed correlation does not 
come as a surprise as the variation of the $R_{23}$ parameter is related 
to the variation of both the ${\rm [OII]\lambda3727/H\alpha}$ ratio, and 
the ionization conditions as traced by the ratio of two different oxygen 
emission lines, i.e., ${\rm 
\log R_{23}\propto~\log([OII]\lambda3727/H\alpha)+\log(1+O_{32})}$, 
where ${\rm O_{32}\,=\,[OIII]\lambda4959,\lambda5007/[OII]\lambda3727}$ 
is an ionization-sensitive ratio. 

The dependence of the abundance-sensitive $R_{23}$ parameter on the 
metallicity is degenerate. Indeed, at a fixed value of $R_{23}$ two 
different values of metallicity are possible: at the same oxygen 
abundance, different ionization parameters lead to different values 
of $R_{23}$ (McCall et al. 1985).  
Different techniques have been developed to break this degeneracy with 
some success (Alloin et al. 1979; McGaugh 1991, van Zee et al. 1998, 
Kobulnicky et al. 1999). 
To estimate the oxygen abundance we have used the calibration of 
McGaugh (1991). This calibration is parameterized as a function of the 
excitation-sensitive parameter ${\rm O_{32}}$. We have used the secondary 
metallicity indicator ${\rm [NII]\lambda6583/H\alpha}$ to determine which 
branch of the McGaugh calibration to use (see Lamareille et al. 2004 for 
a detailed discussion of the abundance estimate for the galaxy sample).
                                                                                
Fig.~\ref{o2ha_oh} shows the relationship between oxygen abundance, 
expressed in terms of ${\rm 12+\log(O/H)}$, and the extinction-corrected 
${\rm [OII]\lambda3727/H\alpha}$ ratio. Large filled circles and the 
associated bars show the means and the standard deviations of the ${\rm 
[OII]\lambda3727/H\alpha}$ ratio distributions in 0.2 dex wide bins.
The relationship between the extinction-corrected ${\rm 
[OII]\lambda3727/H\alpha}$ ratio and oxygen abundance splits into two 
regimes. For metal-poor galaxies, i.e., ${\rm 12+\log(O/H) \la 8.4}$, 
the intrinsic ${\rm [OII]\lambda3727/H\alpha}$ flux ratio increases 
with oxygen abundance. For these galaxies, the Spearman rank correlation 
coefficient is $0.73$, with a two-sided probability of obtaining this 
value by chance almost equal to zero, i.e., $2.5\times\,10^{-51}$. 
This indicates a strong correlation between the extinction-corrected 
${\rm [OII]\lambda3727/H\alpha}$ ratio and oxygen abundance for 
metal-poor galaxies. On the other hand, metal-rich galaxies, i.e., 
${\rm 12+\log(O/H) \ga 8.4}$, show a similar trend but with a slope 
of the opposite sign. The Spearman rank correlation coefficient is 
$-0.6$, with a two-sided probability of obtaining this value by chance 
almost equal to zero, i.e., $2.5\times\,10^{-33}$, suggesting a strong
anti-correlation between the intrinsic ${\rm [OII]\lambda3727/H\alpha}$ 
flux ratio and oxygen abundance. The metal-rich branch in the ${\rm 
[OII]\lambda3727/H\alpha}$ vs. ${\rm 12+\log(O/H)}$ diagram is 
consistent with the same relationship for the NFGS galaxy sample 
constructed by Kewley et al. (2004) using the McGaugh (1991) calibration. 

However, there is a concern here. Because radial abundance gradients are 
known to exist in spiral galaxies, it has been argued, depending on the
metallicity gradients and the relative weight of different {H{\sc ii}}
regions in the integrated emission line spectra, that the ${\rm R_{23}}$
parameter might not be a useful indicator of galaxy overall metallicity
(Stasi\'nska \& Sodr\'e 2001; but see Kobulnicky et al. 1999 for a 
different view).
To establish the dependence of ${\rm [OII]\lambda3727/H\alpha}$ flux 
ratio on the emitting gas metallicity, it is useful to confirm the 
observed correlation in Fig\,\ref{o2ha_oh} using metallicity 
indicators other than the ${\rm R_{23}}$ parameter. 
The ${\rm [NII]\lambda6584/H\alpha}$ ratio has been proposed recently 
as an empirical metallicity indicator (van Zee et al. 1998, 
Denicol\'o et al. 2002). The ${\rm [NII]\lambda6584/H\alpha}$ ratio 
is less sensitive to the electron temperature than the ${\rm R_{23}}$ 
parameter, making this ratio less affected by the doubled value 
problem (e.g., Kewley \& Dopita 2002). A valuable advantage of using 
this emission line ratio is its independence of both reddening and 
the accuracy of the relative flux calibration. In integrated spectra 
of galaxies, one expects however a non negligible contribution from 
a diffuse medium (e.g., Collins et al. 2000; Zurita et al. 2000). 
The ${\rm [NII]\lambda6584/H\alpha}$ ratios in the diffuse medium 
are generally larger than in nearby {H{\sc ii}} regions. The effect 
of the diffuse medium is to increase the ${\rm [NII]\lambda6584/H\alpha}$ 
ratio by about $30\%$ at most: this increase is smaller than the 
metallicity dependence of this ratio, making the ratio a 
useful metallicity indicator (see Stasi\'nska \& Sodr\'e 2001, and 
references therein). It is worth mentioning that our sample galaxies 
are distributed along a well defined sequence in the ${\rm 
[NII]\lambda6584/H\alpha}$ ratio against ${\rm R_{23}}$ 
parameter diagram, interpreted as a metallicity-excitation sequence, 
similar to the sequence defined by local {H{\sc ii}} galaxies (Fig. 14c 
of McCall et al. 1985). Another proposed empirical metallicity indicator 
that does not suffer from the double value problem is the ${\rm 
[NII]\lambda6584/[OII]\lambda3727}$ ratio (Dopita et al. 2000, Kewley 
\& Dopita 2002). It is however strongly dependent on the extinction 
correction, and the spectrophotometric accuracy of the spectra.
 
Fig.\,\ref{o2ha_n2ha} shows the relationship between the 
extinction-corrected ${\rm [OII]\lambda3727/H\alpha}$ ratio and the 
${\rm [NII]\lambda6584/H\alpha}$ ratio. Filled circles and associated 
bars show the means and the standard deviations of the ${\rm 
[OII]\lambda3727/H\alpha}$ ratio in 0.2 dex wide bins. 
For galaxies with ${\rm \log([NII]\lambda6584/H\alpha)\ga\,-1}$, 
corresponding roughly to ${\rm \log(O/H)+12\ga\,8.4}$ using the 
van Zee et al. (1998) calibration, the
Spearman correlation coefficient is $-0.36$, with the two-sided 
probability of obtaining this value by chance of $9\,\times\,10^{-27}$.
This indicates a statistically significant anti-correlation between 
the metallicity indicator and the extinction-corrected ${\rm 
[OII]\lambda3727/H\alpha}$ ratio. For metal-rich galaxies, i.e., 
${\rm \log(O/H)+12\ga\,8.4}$, nitrogen is thought to be predominantly 
a secondary element (e.g., Villa-Costas \& Edmunds 1993, Henry et al. 
2000), so the observed trend reflects the sensitivity of the intrinsic
${\rm [OII]\lambda3727/H\alpha}$ ratio to abundance within this 
metallicity regime. 
For galaxies with low ${\rm [NII]\lambda6584/H\alpha}$ ratio, i.e., 
mainly metal-poor galaxies for which nitrogen is a primary element 
(Matteucci 1986), the relationship between ${\rm 
[NII]\lambda6584/H\alpha}$ and the extinction-corrected ${\rm 
[OII]\lambda3727/H\alpha}$ flux ratio is reversed. The Spearman 
correlation coefficient is $0.33$, with the two-sided probability 
of obtaining this value by chance of $1.3\,\times\,10^{-8}$. 
This indicates a statistically significant correlation between 
the metallicity indicator and the extinction-corrected ${\rm 
[OII]\lambda3727/H\alpha}$ flux ratio for metal-poor galaxies. 
This confirms that the variation of extinction-corrected ${\rm 
[OII]\lambda3727/H\alpha}$ ratio is coupled with the evolution 
of metallicity.


\subsection{ ${\rm [OII]\lambda3727/H\alpha}$ ratio and excitation 
             state }
\label{o2ha_excit}

Kewley et al. (2004) have found that for ${\rm 12+log(O/H)\,\ga\,8.5}$,  
the variation of extinction-corrected ${\rm [OII]\lambda3727/H\alpha}$ 
flux ratio does not depend on the ionization state of interstellar 
emitting gas. 
For our galaxy sample, the scatter of the extinction-corrected ${\rm 
[OII]\lambda3727/H\alpha}$ flux ratio at a given metallicity appears 
to be related to the variation of the ionization parameter in galaxies. 
Indeed, at a given oxygen abundance, galaxies with large 
ionization-sensitive ${\rm [OIII]\lambda5007/H\beta}$ ratio, shown as 
filled squares in Fig.~\ref{o2ha_oh}, tend to have lower intrinsic 
${\rm [OII]\lambda3727/H\alpha}$ flux ratio than galaxies with a 
low-to-intermediate ionization-sensitive ratio. This suggests that the 
variation of the excitation state of the interstellar emitting gas in 
galaxies may contribute to the observed variation of the ${\rm 
[OII]\lambda3727/H\alpha}$ ratio.

\begin{figure*}
\includegraphics[clip=,width=0.45\textwidth]{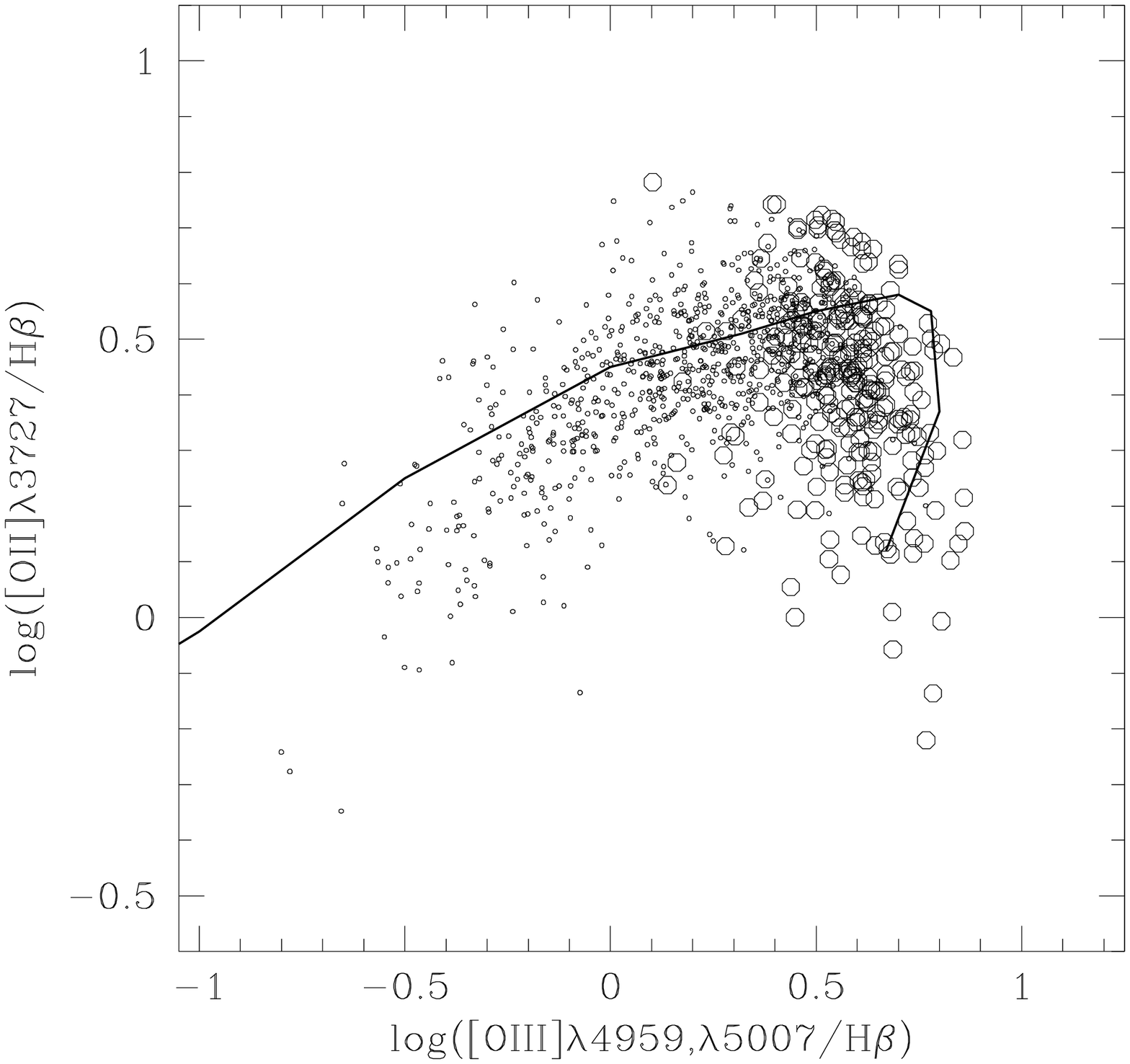}
\includegraphics[clip=,width=0.45\textwidth]{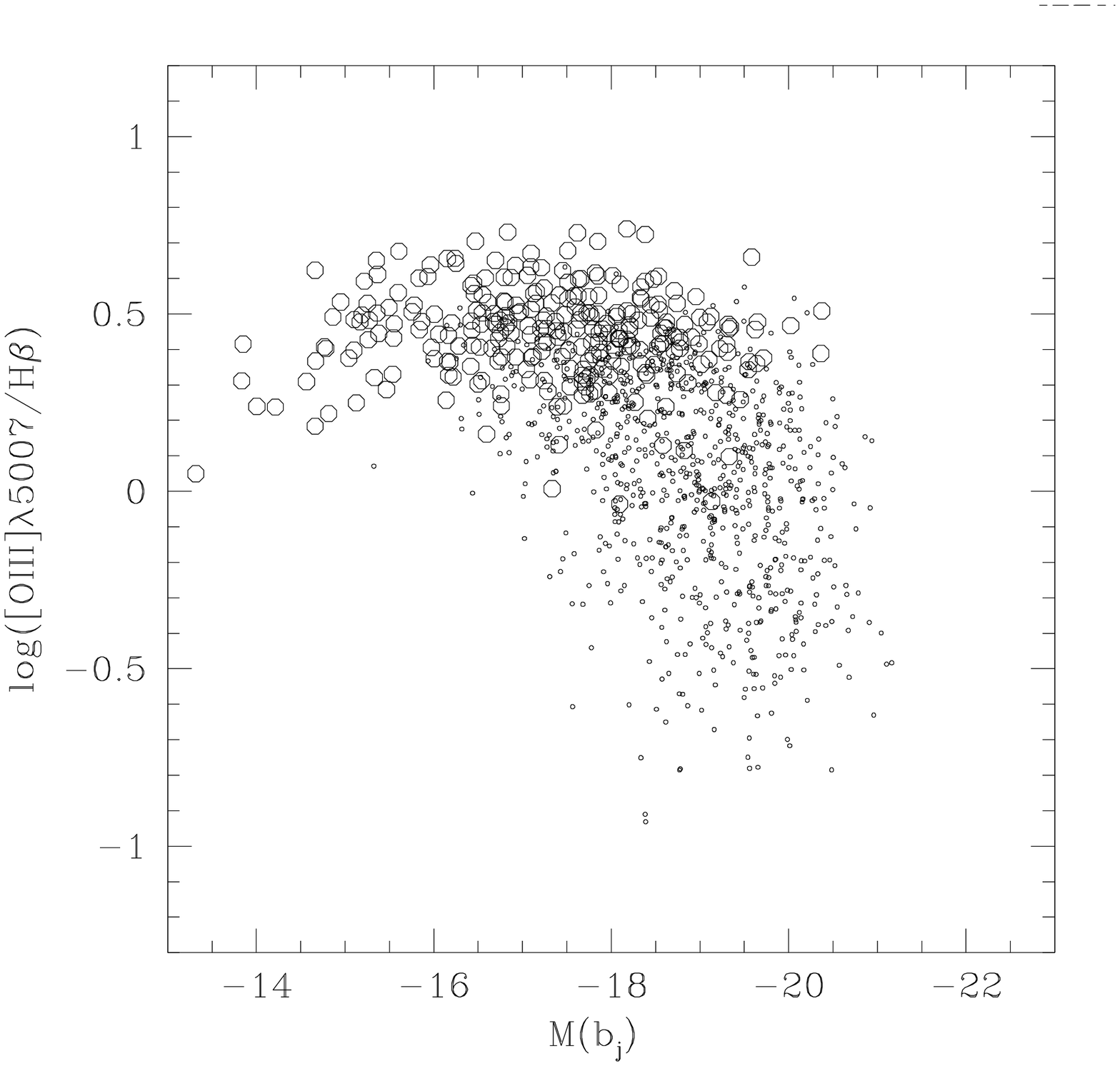}
\caption{Left: The relationship between ${\rm [OII]\lambda3727/H\beta}$ 
and ${\rm [OIII]\lambda4959,\lambda5007/H\beta}$ for our sample galaxies. 
Large (small) circles show metal-poor (metal-rich) galaxies 
($12+\log(O/H)\le 8.4 \, [> 8.4]$). The solid line shows the theoretical 
sequence from McCall et al. (1985), which fit the local HII galaxies 
with metallicity decreasing from the left to the right. Right: Excitation 
sensitive ratio versus $b_j$-band absolute magnitude for our sample 
galaxies.}
\label{o2hb_o3hb_Mbj}
\end{figure*}

\begin{figure}
\includegraphics[clip=,width=0.45\textwidth]{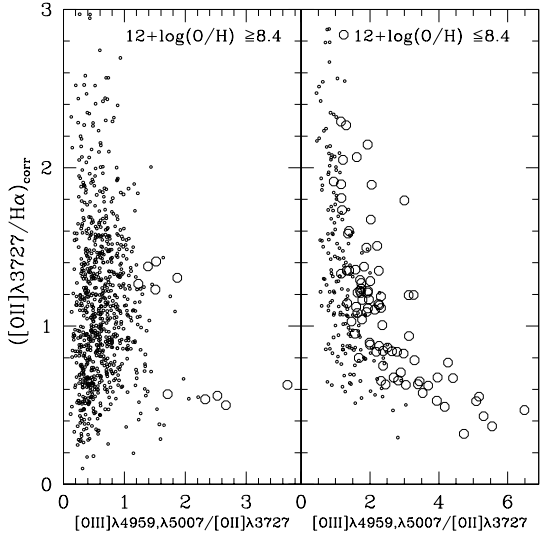}
\caption{Left: the relationship between ionization parameter 
sensitive ${\rm [OIII]\lambda4959\lambda5007/[OII]\lambda3727}$ 
and the ${[OII]\lambda3727/H\alpha}$ ratio for metal-rich galaxies 
($12+\log(O/H)\ge 8.4$). Right: similar to the left panel but for 
metal-poor galaxies ($12+\log(O/H)\le 8.4$). Large circles show 
galaxies with ${\rm \log([OIII]\lambda5007/H\beta)\ge 0.5}$.   }
\label{o2ha_o32}
\end{figure}

The left panel of Fig.~\ref{o2hb_o3hb_Mbj} shows the diagnostic 
diagram of ${\rm [OII]\lambda3727/H\beta}$ ratio as a function of 
${\rm [OIII]\lambda4959,\lambda5007/H\beta}$ for our galaxy sample. 
Large/small circles show metal-poor/metal-rich galaxies, i.e., 
${\rm 12+\log(O/H)\le 8.4 (> 8.4)}$.
The continuous line shows the theoretical sequence of McCall, Rybski,
\& Shields (1985) for line ratios of {H{\sc ii}} galaxies as a function 
of metallicity. Along the track, the metallicity is high at the lower 
left, i.e., for low excitation systems, and low at the upper right, i.e., 
for high excitation systems (McCall et al. 1985). Most of the metal-poor 
galaxies in the sample lie in the moderate- to high-excitation regime 
populated by local {H{\sc ii}} galaxies, i.e., ${\rm 
\log([OIII]\lambda5007/H\beta)\ge 0.3}$, while metal-rich galaxies 
are located in the low-excitation regime. The right panel of 
Fig.\,\ref{o2hb_o3hb_Mbj} shows the ${\rm [OIII]\lambda5007/H\beta}$ 
ratio versus the absolute $b_{j}$-band magnitude for our sample 
galaxies. The galaxies define a continuous sequence in this 
diagram. The observed sequence is interpreted as being a variation 
in the metallicity of the ionized gas (Dopita \& Evans 1986; 
Stasi\'nska 1990). On average faint/metal-poor galaxies tend to be 
highly ionized, while bright/metal-rich galaxies are characterized 
by low-ionization parameters. 

The line ratio ${\rm O_{32}}$ is a function of both ionization 
parameter and metallicity (Kewley \& Dopita 2002). For a galaxy sample 
that covers a large range of metallicity, a given ${\rm O_{32}}$ could 
correspond to different combinations of abundances and ionization 
parameters. In order to distinguish between the effects of ionization 
and metallicity, we have split the galaxy sample into metal-rich, i.e., 
${\rm 12+\log(O/H)>\,8.4}$, and metal-poor, i.e., ${\rm 
12+\log(O/H)\,\le\,8.4}$, galaxy subsamples. Fig.~\ref{o2ha_o32} shows 
the extinction-corrected ${\rm [OII]\lambda3727/H\alpha}$ flux ratio 
versus ${\rm O_{32}}$ ratio for metal-rich and metal-poor subsamples 
respectively. The ${\rm O_{32}}$ ratio has been estimated using emission 
line equivalent widths. Kobulnicky \& Phillips (2003) have shown that 
estimates of this ratio using equivalent widths give results similar 
to using emission line fluxes. Large circles show galaxies with ${\rm 
\log([OIII]\lambda5007/H\beta)\ge\,0.5}$. Note that for the metallicity 
range covered by galaxies in our sample, the ${\rm 
[OIII]\lambda5007/H\beta}$ ratio is sensitive mainly to ionization 
parameter, and is almost independent of metallicity (Kewley et al. 2004).  

For metal-rich galaxies, the Spearman correlation coefficient for the 
relationship between extinction-corrected ${\rm [OII]\lambda3727/H\alpha}$ 
ratio and ${\rm O_{32}}$ ratio is $0.07$, with the two-sided probability 
of obtaining this value by chance of 0.05. This indicates that there is
no statistically significant correlation between the intrinsic ${\rm 
[OII]\lambda3727/H\alpha}$ flux ratio and the ionization-sensitive 
${\rm O_{32}}$ ratio, in agreement with the Kewley et al. (2004) result. 
The subsample of metal-rich galaxies spans a limited range in 
ionization-sensitive ratio; the distribution of ${\rm O_{32}}$ ratio 
for metal-rich galaxies has a mean of $0.63$, with an accuracy of $0.01$, 
and $70\%$ of the galaxies of this subsample with a ratio less than $0.5$.
The low ${\rm O_{32}}$ ratio suggests that for metal-rich galaxies, a 
significant fraction of oxygen emission results from ${\rm O^{+}}$ 
species.

Metal-poor galaxies exhibit a larger range of ionization-sensitive
diagnostic ratios, extending to extreme excitation states. The majority 
of galaxies in our sample with large excitation-sensitive ratios are 
metal-poor. The distribution of the ${\rm O_{32}}$ ratio for metal-poor 
galaxies has a mean of $1.66\pm 0.07$, with $72\%$ of the galaxies  
having ${\rm O_{32}}$ larger than unity. 

Metal-poor galaxies with low to moderate excitation, i.e., 
${\rm \log([OIII]\lambda5007/H\beta)\,<\, 0.5}$ cover a wide range in 
ionization-sensitive diagnostic ratios, i.e., ${\rm 0.2\la O_{32}\la 3}$, 
with a mean of $1.23\pm 0.04$. For these galaxies, the Spearman 
correlation coefficient for the relationship between extinction-corrected 
${\rm [OII]\lambda3727/H\alpha}$ ratio and ${\rm O_{32}}$ ratio is 
$-0.67$, with a two-sided probability of obtaining this value by chance 
of $1.1\times\,10^{-25}$. This indicates a statistically significant 
anti-correlation between the intrinsic ${\rm [OII]\lambda3727/H\alpha}$ 
flux ratio and the ionization-sensitive ${\rm O_{32}}$ ratio. 
Highly ionized metal-poor galaxies, i.e., ${\rm 
\log([OIII]\lambda5007/H\beta)\ge 0.5}$, span a large range of ionization
parameter, i.e., ${\rm 1\la O_{32}\la 10}$, with a mean ratio of 
$2.59\pm 0.16$. These galaxies show a stronger anticorrelation between 
the extinction-corrected ${\rm [OII]\lambda3727/H\alpha}$ flux ratio and 
the ionization-sensitive ${\rm O_{32}}$ ratio: the Spearman correlation 
coefficient for these galaxies is $-0.8$, with the two-sided probability 
of obtaining this value by chance of $2.4\times\,10^{-18}$. 

\begin{figure*}
\includegraphics[clip=,width=0.4\textwidth]{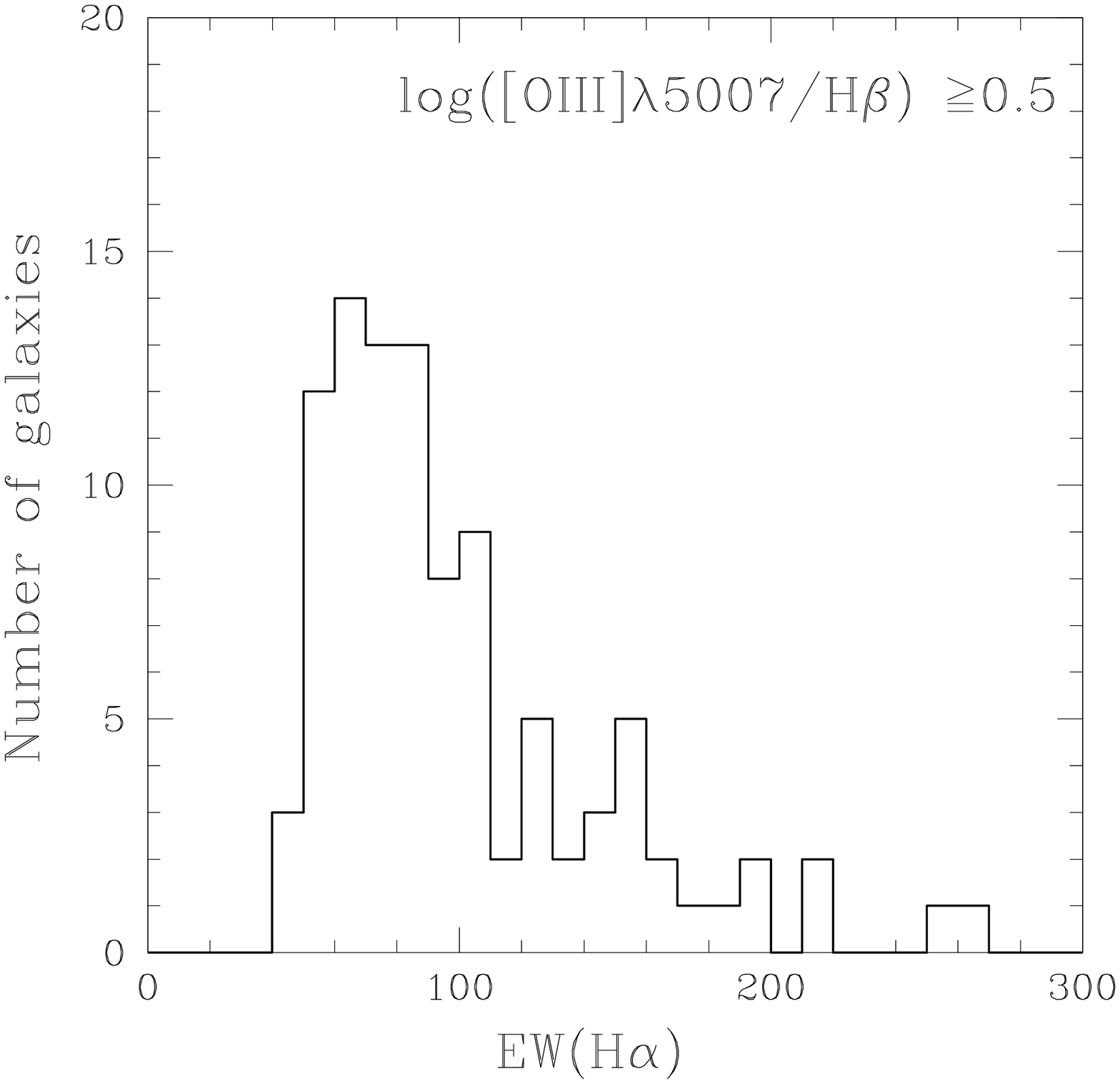}
\includegraphics[clip=,width=0.4\textwidth]{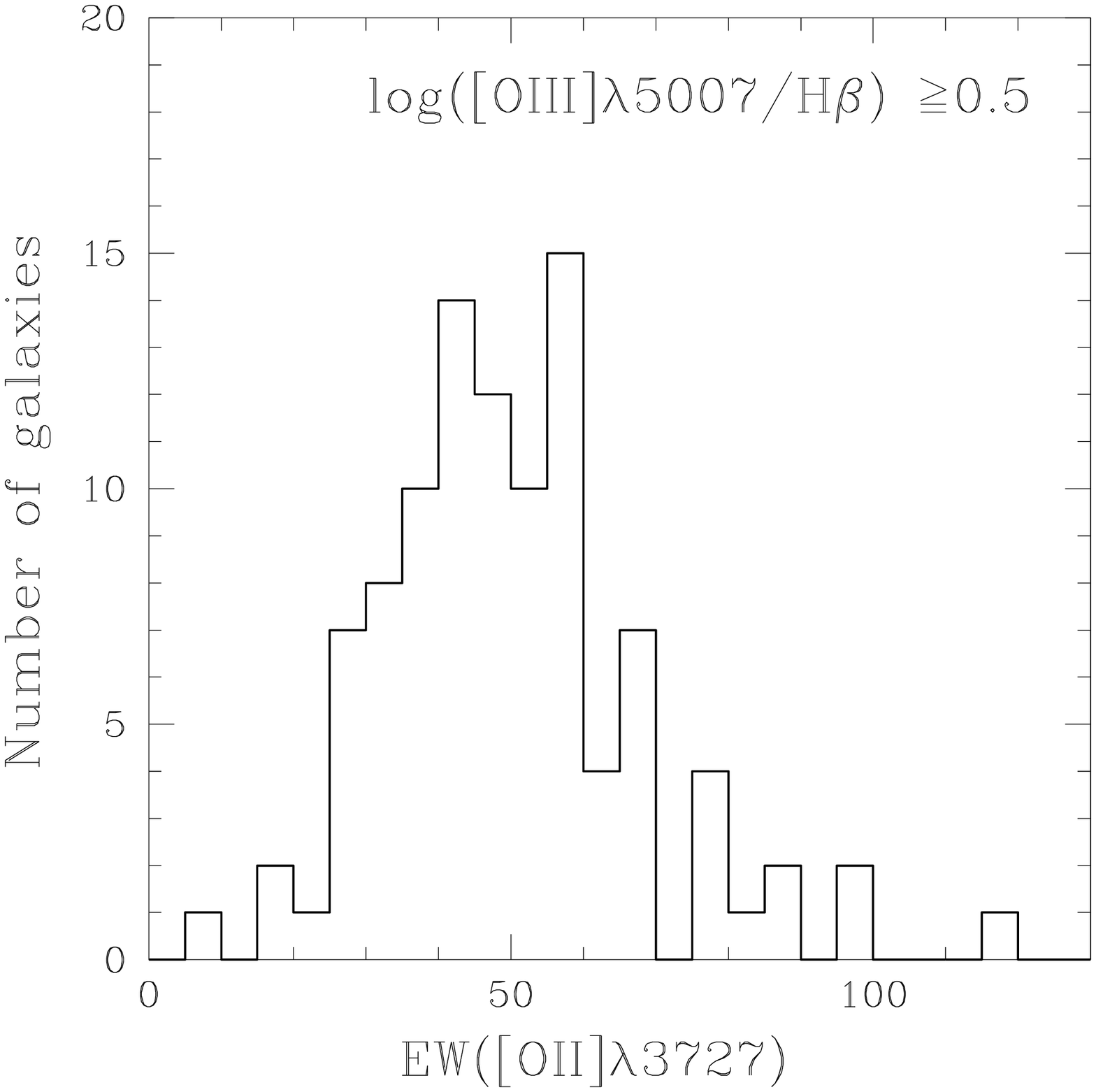}
\includegraphics[clip=,width=0.4\textwidth]{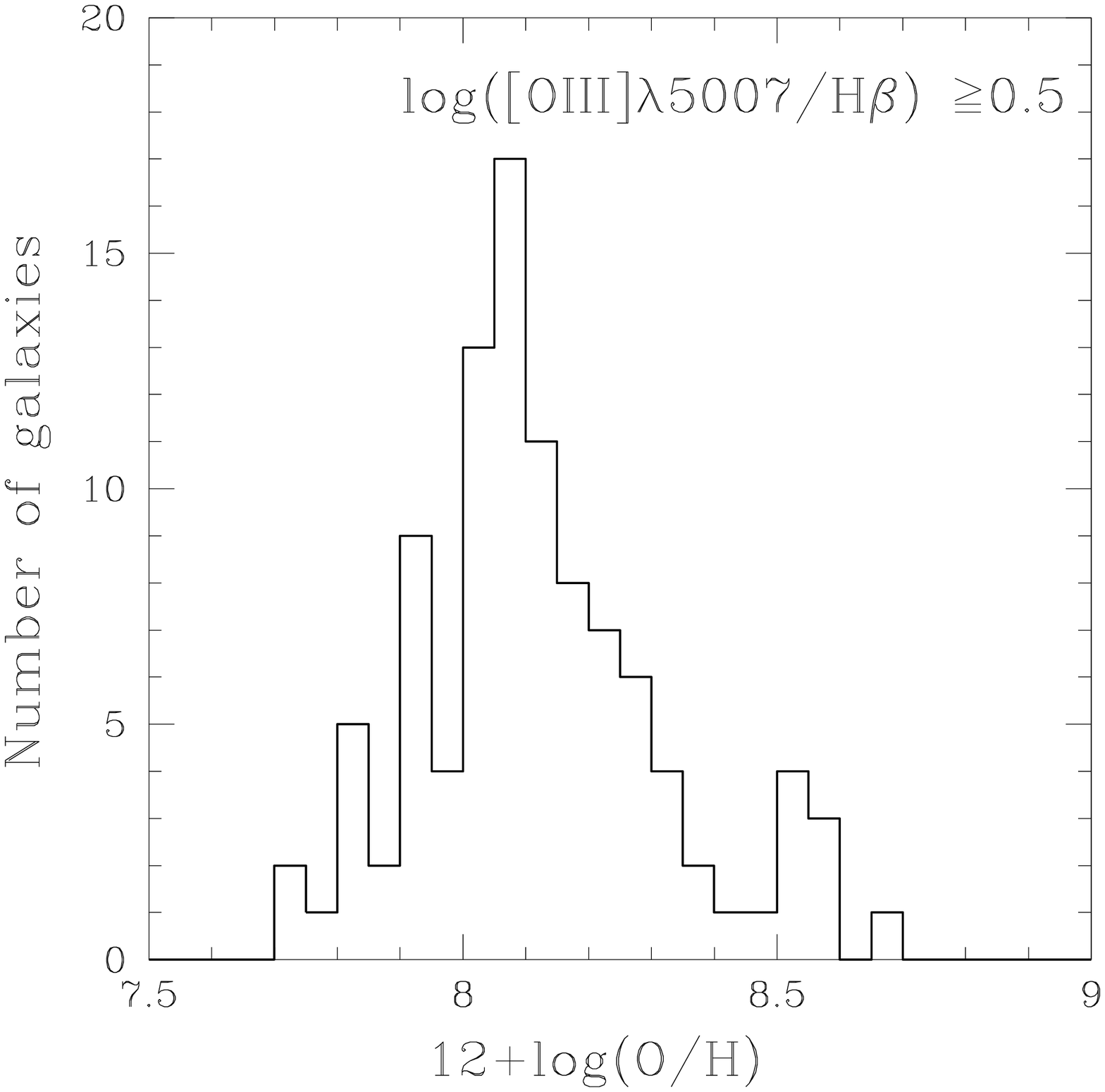}
\includegraphics[clip=,width=0.4\textwidth]{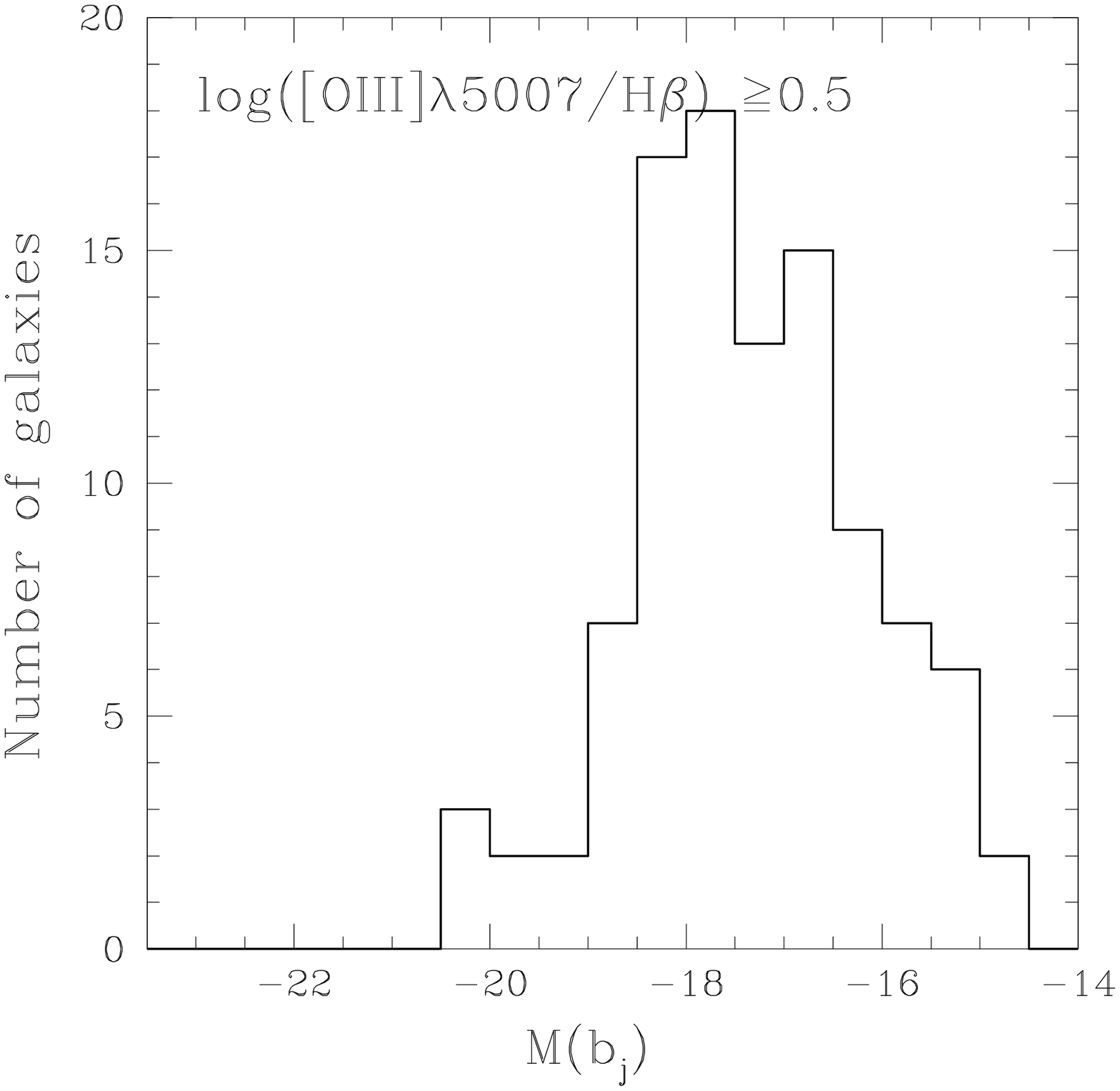}
\caption{Distribution of properties of highly ionized galaxies, i.e., 
${\rm \log([OIII]\lambda5007/H\beta)\,\ge\, 0.5}$. The upper panels 
shows the equivalent widths of ${\rm [OII]\lambda3727}$ and ${\rm 
H\alpha}$ emission lines, with medians of 85\,{\AA} and 48\,{\AA} 
respectively. The lower panels show the distributions of oxygen 
abundance and absolute $b_j$-band magnitude. The median of oxygen 
abundance distribution is ${\rm 12+\log(OH)\,=\,8.1 (\sim\,Z_{\odot}/4)}$, 
and $-17.4$ for $b_j$-band magnitude distribution.}
\label{dist_highionz}
\end{figure*}

The lack of a dependence of the extinction-corrected ${\rm 
[OII]\lambda3727/H\alpha}$ ratio on ionization state of the interstellar 
medium for NFGS galaxies may be attributed to the absence of such highly 
ionized metal-poor galaxies. This galaxy sample consists mostly of normal 
star-forming galaxies, with few active starburst galaxies and extremely 
metal-poor dwarfs (Jansen et al. 2000). The observed dependence 
of extinction-corrected ${\rm [OII]\lambda3727/H\alpha}$ flux ratio on  
ionization state of the interstellar medium for metal-poor and highly 
ionized galaxies suggests that at low metallicity, where the electronic 
temperature is very high, a significant fraction of oxygen atoms may be 
in the form of ${\rm O^{++}}$ and higher excitation levels.  
For metal-poor galaxies with a high ionization parameter, the variation 
of the extinction-corrected ${\rm [OII]\lambda3727/H\alpha}$ ratio 
is regulated by the variation of the ionization parameter rather than 
metallicity. Fig.\,\ref{dist_highionz} shows the distributions of 
${\rm H\alpha}$ and ${\rm [OII]\lambda3727}$ emission line equivalent 
widths, oxygen abundance, and $b_j$-band absolute magnitude for galaxies 
with ${\rm \log([OIII]\lambda5007/H\beta)\ge 0.5}$. The highly ionized 
galaxy population in our sample, for which a strong anti-correlation is
observed between the extinction-corrected ${\rm [OII]\lambda3727/H\alpha}$
flux ratio and the excitation-sensitive ratio ${\rm O_{32}}$, consists 
mainly of faint metal-poor galaxies, in which the starburst is still 
vigorously active, keeping the high ionization conditions. 
An important conclusion regarding the high ionization galaxy population
is that the estimate of their star formation rate based on the ${\rm
[OII]\lambda3727}$ luminosity may be significantly underestimated, even 
when the dependence of the extinction-corrected ${\rm 
[OII]\lambda3727/H\alpha}$ ratio on metallicity is corrected. Guzm\'an 
et al. (1997) have shown that for a $z = 0.137$ compact field galaxy 
with extreme ionization-sensitive ratios, i.e., ${\rm O_{32}=3.8}$ and 
${\rm [OIII]\lambda5007/H\beta=5.27}$, the star formation rate based 
on ${\rm [OII]\lambda3727}$ luminosity is underestimated by a factor 
of $6$, compared to the star formation rate based on ${\rm H\alpha}$ 
luminosity.

\section{Summary and conclusions}
\label{concl.sec}

We have used spectrophotometric data for a sample of 1\,124 nearby 
star-forming galaxies from the 2dFGRS sample, spanning a range of $7$ 
magnitudes in ${\rm M(b_j)}$, to investigate the systematic variation 
of the ${\rm [OII]\lambda3727/H\alpha}$ emission-line ratio as a 
function of galaxy and interstellar emitting gas properties. 

The 2dF fibres cover, on average, about $11\%$ of the total light of 
the galaxy. No evidence for systematic aperture bias affecting the 
estimate of the emission line properties is found. This suggests that 
our spectra are sufficiently representative of the integrated galaxy 
spectra.
The nebular extinction as derived from the Balmer decrement is found 
to correlate with the intrinsic absolute luminosity. The mean of the 
distribution of the extinction-corrected emission line ${\rm 
[OII]\lambda3727/H\alpha}$ flux ratio is similar to what was found 
for other galaxy samples, selected in different ways, confirming that 
the internal reddening is a driver behind the variation of the observed 
${\rm [OII]\lambda3727/H\alpha}$. 

We confirm that there is a strong correlation between the
extinction-corrected ${\rm [OII]\lambda3727/H\alpha}$ ratio and the
oxygen abundance for metal-rich galaxies, and extend the observed
correlation further to the metal-poor regime, i.e., ${\rm 
12+\log(O/H)\la\,8.4}$. This relationship consists of two branches, 
i.e., where the ${\rm [OII]\lambda3727/H\alpha}$ ratio is increasing 
(decreasing) as a function of the oxygen abundance for ${\rm 
12+\log(O/H)\la 8.4(\ga 8.4)}$. For metal-rich galaxies, there is no 
clear dependence of the extinction-corrected ${\rm 
[OII]\lambda3727/H\alpha}$ ratio on the ionization parameter, in 
agreement with what was reported for NFGS sample galaxies. However, a 
strong correlation is seen for metal-poor galaxies, especially for those 
with high ionization-sensitive ratios. These galaxies tend to be faint 
and strong ${\rm [OII]\lambda3727}$ emitters. For these galaxies, the 
${\rm [OII]\lambda3727/H\alpha}$ ratio is more sensitive to the variation 
of ionization parameter than to the variation of oxygen abundance. 

An emission-line galaxy spectrum is the result of many physical 
properties of the ionized gas, e.g., the chemical abundance and dust 
content, and of the relative importance of the ongoing star-forming 
activity, e.g., the star formation timescale. The excitation state 
depends both on the emitting gas abundance and on the ionizing stellar 
flux, which in turn depends on the effective temperature of the ionizing 
stars, which depends on the stellar initial metallicity, and on the 
age of the ongoing star formation event. Different detection techniques 
preferentially detect emission line galaxies at different stages of the 
starburst. An important conclusion is that using the ${\rm 
[OII]\lambda3727}$ emission line as a star formation rate indicator 
requires a good understanding of the selection criteria of the galaxy 
sample under investigation, and how they determine its properties, i.e., 
extinction, metallicity, and excitation state.

\section*{Acknowledgements}
M. M. would like to thank warmly A. Arag\'on-Salamanca for highlighting 
discussions, and sharing his results prior to publication. We thank an 
anonymous referee for useful comments which significantly improved our 
paper.

\bsp

\label{lastpage}

\end{document}